\documentclass[twocolumn]{aastex631}
\usepackage{bm}
\usepackage{amsmath,amstext}

\shorttitle{Excitation of Turbulence and Oscillation in Flare Loop Top}
\shortauthors{Shibata et al.}
\graphicspath{{./}{figures/}}

\begin{document}

\title{Numerical Study on Excitation of Turbulence and Oscillation \\
in Above-the-loop-top Region of a Solar Flare}

\author{Kengo Shibata}
\affiliation{Department of Earth and Space Science, Graduate School of Science, Osaka University, Toyonaka, Osaka 560-0043, Japan}

\author[0000-0003-3882-3945]{Shinsuke Takasao}
\affiliation{Department of Earth and Space Science, Graduate School of Science, Osaka University, Toyonaka, Osaka 560-0043, Japan}

\author[0000-0002-6903-6832]{Katharine K. Reeves}
\affiliation{Harvard-Smithsonian Center for Astrophysics, 60 Garden Street, Cambridge, MA 02138, USA}



\begin{abstract}
Extreme ultraviolet imaging spectroscopic observations often show an increase in line width around the loop-top or above-loop-top (ALT) region of solar flares, suggestive of turbulence. In addition, recent spectroscopic observations found the oscillation in the Doppler velocity around the ALT region. We performed three-dimensional magnetohydrodynamic (MHD) simulations to investigate the dynamics in the ALT region, with a particular focus on the generation of turbulence and the excitation of the oscillatory motion. We found a rapid growth of MHD instabilities around the upper parts of the ALT region (arms of the magnetic tuning fork). The instabilities grow more rapidly than the magnetic Rayleigh-Taylor-type instabilities at the density interface beneath the reconnecting current sheet. Eventually, the ALT region is filled with turbulent flows. The arms of the magnetic tuning fork have bad-curvature and transonic flows. Therefore, we consider that the rapidly growing instabilities are combinations of pressure-driven and centrifugally driven Rayleigh-Taylor-type instabilities. Despite the presence of turbulent flows, the ALT region shows a coherent oscillation driven by the backflow of the reconnection jet. We examine the numerical results by re-analyzing the solar flare presented in \citet{Reeves2020ApJ}. We find that the highest non-thermal velocity is always at the uppermost visible edge of the ALT region, where oscillations are present. This result is consistent with our models.
We also argue that the turbulent magnetic field has a significant impact on the confinement of non-thermal electrons in the ALT region.
\end{abstract}



\section{Introduction} \label{sec:intro}
Magnetic reconnection is the central mechanism that powers solar flares by suddenly releasing magnetic energy \citep[see reviews by, e.g., ][]{Shibata2011LRSP,Fletcher2011SSRv,Hudson2011SSRv}. 
Solar flares are suitable targets to study the energy conversion process of magnetic reconnection in astrophysical systems.
Solar flares form X-ray bright flare loops with a temperature of $\gtrsim 10^7$~K. The soft X-ray structures of the flare loops are often accompanied by localized hard X-ray sources \citep[e.g.][]{Holman2011SSRv}. 
The hard X-ray sources that appear just above the soft X-ray flare loops are called loop-top or above-the-loop-top (ALT) sources \citep[e.g.][]{Masuda1994Natur,Petrosian2002ApJ,Liu2013ApJ}. The meaning of "above" is not strict because of the limitations of spatial resolution and the line-of-sight effect, and we hereafter use the term of ALT. Recent radio observations suggested a concentration of nonthermal electrons around the ALT region \citep{BinChen2020NatAs}.
Such observations have motivated theoretical studies to understand the connection between the magnetohydrodynamic (MHD)-scale and the kinetic-scale processes.

Although the electron acceleration occurs essentially at kinetic scales, MHD-scale dynamics should also be considered to understand the origin of the ALT sources in hard X-rays.
Super-magnetosonic reconnection outflows will produce termination shocks in the ALT regions. 
Theoretical studies suggest an efficient production of nonthermal electrons at the termination shock \citep[e.g.][]{Tsuneta1998ApJ,Nishizuka2013PhRvL,Kong2019ApJ,Kong2020ApJ}.
Electrons could also be efficiently accelerated in the contracting magnetic loops via magnetic mirror, which may produce an ALT source \citep{Somov1997ApJ}.
If MHD turbulence develops, the stochastic electron acceleration will take place as a result of the energy cascade down to the kinetic scale \citep{Petrosian2006ApJ,Petrosian2012SSRv}.
In any cases, the MHD-scale structures such as shocks, turbulence, and magnetic mirrors determine which processes operate effectively. Therefore, it is important to reveal the detailed MHD structure around the ALT regions.

\citet{Takasao2016ApJ} investigated the dynamic behavior of the ALT region using 2D MHD simulations \citep[see also][]{Takasao2015ApJ}. It is found that the reconnection outflow impinging on the reconnected loops forms multiple termination shocks and excites local oscillation in the ALT region, even when the reconnection outflow is laminar and quasi-steady. As the magnetic structure in the oscillating ALT region is similar to a tuning fork, they termed it ``magnetic tuning fork."
Later, more sophisticated 2D simulations of plasmoid-mediated reconnection have been performed to find that the ALT oscillation can also occur in such cases but in an asymmetric manner \citep{Takahashi2017ApJ,Shen2018ApJ}, suggesting the robustness of the magnetic tuning fork mechanism \citep[for the case in partially ionized plasmas, see][]{Murtas2022PhPl}.
Spectroscopic observations of a solar flare with IRIS and EIS/Hinode provided supporting evidence; \citet{Reeves2020ApJ} identified oscillating plasma motions in the loop-top region of an X-class flare using Doppler shift measurements.

Apart from the oscillation, observations show indications of turbulent plasma motions around the ALT regions.
Turbulent structures are discerned in extreme ultra-violet images \citep[e.g.][]{McKenzie2013ApJ,Shen2022NatAs, Freed2018ApJ}, and spectroscopic observations also indicate local enhancement of nonthermal line widths \citep[e.g.][]{Hara2008PASJ,Doscheck2014ApJ,Warren2018ApJ,Reeves2020ApJ}. From coordinated observations, \citet{Kontar2017PhRvL} argued that the kinetic energy of turbulent motions around the loop top seems to be significant in terms of the nonthermal electron energy. This suggests the importance of turbulence for electron acceleration, although the relation with the kinetic-scale process remains unclear \citep[c.f.][]{Petrosian2006ApJ}.

The spatial distribution of turbulence can change the scenario of electron acceleration in the ALT regions. If the turbulence develops just around the termination shocks, diffusive shock acceleration will work because of a cross-field diffusion of electrons \citep{Kong2019ApJ,Kong2020ApJ}. However, the turbulent regions may be separated from the shocks. \citet{Shen2022NatAs} performed 3D simulations of a solar flare and examined the development of turbulent flows around the ALT regions \citep[see also][]{Guo2014ApJ,Innes2014ApJ}. They found the development of turbulence beneath the reconnection current sheet as a result of instabilities, but the turbulent region is distant from the termination shocks. Such turbulent flows will have a small impact on electron acceleration and confinement.

We examine the excitation and the spatial distribution of turbulent flows in the ALT region using 3D MHD simulations. Our models indicate the development of turbulent flows beneath the current sheet, as shown in previous studies. In addition, we found that turbulent flows develop more rapidly in the arms of the magnetic tuning fork. This study indicates that the ALT region should be full of turbulent flows with multiple shocks. Section~\ref{sec:method} describes the numerical method. The turbulent ALT structure is analyzed in Section~\ref{sec:results}. We summarize the results and give some theoretical discussions on the growth of the instability in Section~\ref{sec:summary_discussion}.

\section{Numerical Setup} \label{sec:method}
\subsection{Basic Equations}

We show 2D and 3D simulations of a solar flare to highlight the importance of three-dimensionality. The MHD equations in the following form are solved:
\begin{align}
    \frac{\partial \rho}{\partial t} + \nabla \cdot (\rho \bm{v})=0,\label{1}\\
    \frac{\partial}{\partial t}(\rho \bm{v})+\nabla \cdot \left[ \rho \bm{vv}+\left( p+\frac{B^2}{8\pi}\right)\bm{I}-\frac{\bm{BB}}{4\pi} \right] =0,\label{2}\\
    \frac{\partial \bm{B}}{\partial t} + c\nabla \times \bm{E} = 0,\label{3}\\
    \frac{\partial}{\partial t}\left( \frac{p}{\gamma-1} + \frac{1}{2}\rho v^2 + \frac{B^2}{8\pi}\right)\nonumber \\
    + \nabla \cdot  \left(\frac{\gamma}{\gamma-1}p + \frac{1}{2}\rho v^2 + \frac{c}{4\pi}\bm{E} \times \bm{B} \right)=0,\label{4}\\
    \bm{E} = \eta \bm{J} - \frac{1}{c}\bm{v} \times \bm{B},\label{5} \\
    \bm{J} = \frac{c}{4\pi}\nabla \times \bm{B}, \label{6}\\
    p = \frac{\rho R T}{\mu}, \label{7}
\end{align}
where $\rho, p,$ and $\bm{v}$ represent the density, the pressure, and the velocity of the gas, respectively. $\bm{E}, \bm{B}$ are the electric and magnetic fields, respectively. $\bm{J}$ is the current density. $\eta$ is the electric resistivity. $\gamma=5/3$ is the specific heat ratio. $R, \mu$, and $T$ are the gas constant, average molecular weight, and the temperature of the gas, respectively. $\bm{I}$ is the unit tensor. Radiative cooling, heat conduction, and gravity are ignored for simplicity. However, we present a 3D simulation including the heat conduction but with a lower spatial resolution in Appendix~A and argue that the main findings of this study are less affected by heat conduction.

We used Athena++ \citep{Stone2020ApJS} to numerically solve the basic equations.
We adopted the Harten-Lax-van Leer Discontinuities (HLLD) approximate Riemann solver \citep{Miyoshi2005JCoPh} and the constrained transport method \citep{Stone2009NewA} to integrate the equations. The piece-wise parabolic method (PPM) is used for spatial reconstruction, and the third-order Runge-Kutta time integration is performed. 
The normalization units of our simulations are summarized in Table \ref{table:1}.

\begin{table}[h]
  \caption{Normalization Units}
  \label{table:1}
  \centering
  \begin{tabular}{lll} \hline
    Quantity & Unit & Value \\ \hline \hline
    Length & $L_{0}$  & $3,000\ \rm{[km]}$\\ 
    Density & $\rho_{0}$  & $1.6 \times 10^{-15} \ \rm{[g\ cm^{-3}]}$\\ 
    Temperature & $T_{0}$  & $2.0 \times 10^{6} \ \rm{[K]}$\\ 
    Velocity & $c_{\rm{iso,0}} = \sqrt{\frac{RT_{0}}{\mu}}$ & $170 \ \rm{[km\ s^{-1}]}$ \\
    Time & $t_{0}= L_{0}/c_{\rm{iso,0}}$  & $17.6 \ \rm{[s]}$\\ 
    Pressure & $p_{0}=\rho_{0}c_{\rm{iso,0}}^2$ & $0.47\  \rm{[erg\ cm^{-3}]}$ \\ 
    Magnetic field & $B_{0}=\sqrt{4\pi \rho_{0}} c_{\rm{iso,0}}$ & $2.4\ [\rm{G}]$ \\ \hline
  \end{tabular}
\end{table}


\subsection{Initial and Boundary Conditions}
\begin{figure}
    \label{figure:1}
    \centering
    \includegraphics[width=1.0\columnwidth]{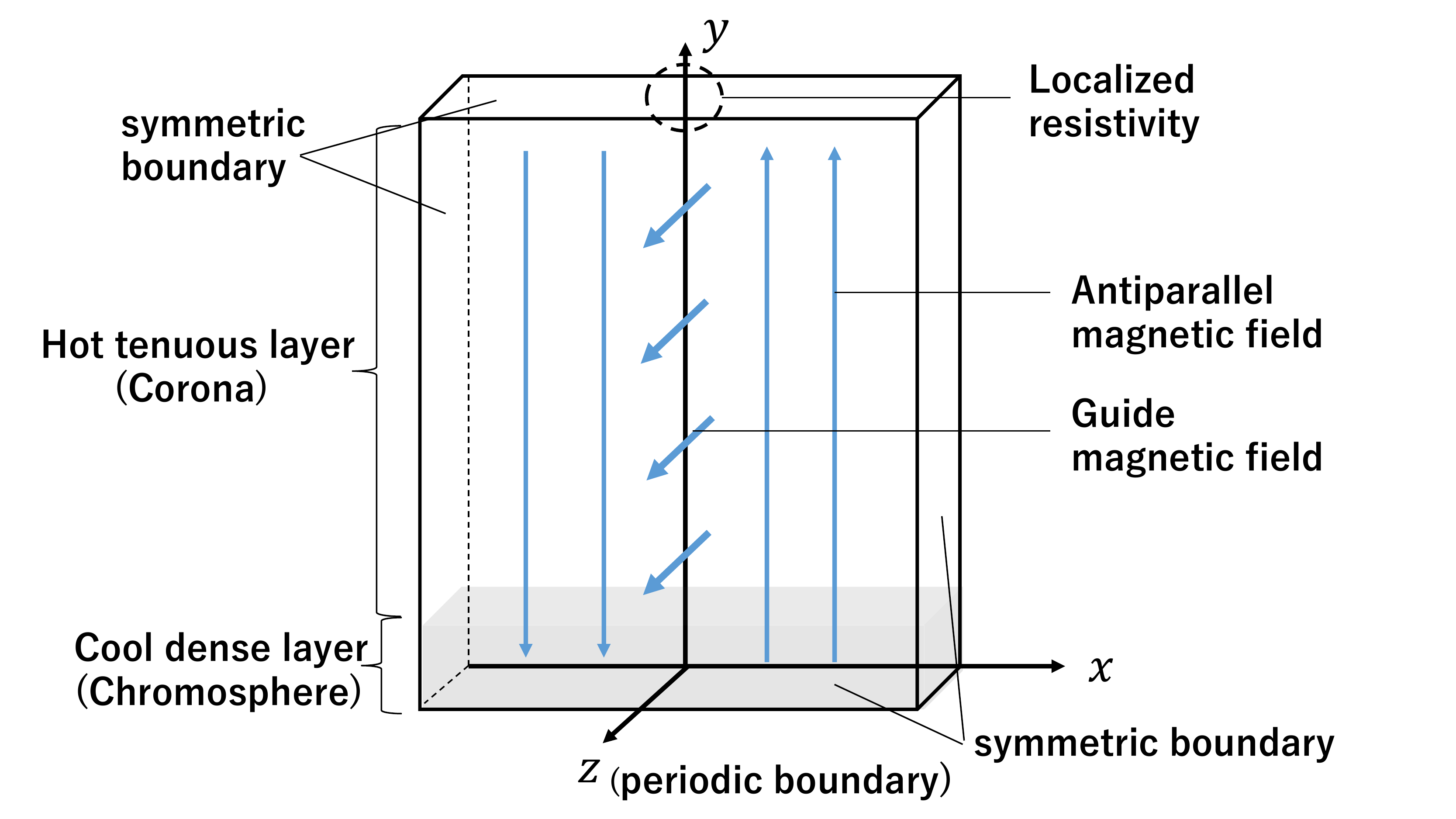}
    \caption{Schematic diagram of the initial and boundary conditions of the 3D
MHD simulation.}\label{fig:ic}
\end{figure}

The atmospheric structure and magnetic field geometry at the initial condition are shown in Figure~\ref{fig:ic}.
Our 3D models are based on \cite{Takasao2015ApJ} and \cite{Takasao2016ApJ}, but the domain size and dimension are different. The calculation domain covers $-7.5L_{0} \leq x \leq 7.5L_{0},\ 0 \leq y \leq 20L_{0},$\ and $-0.75L_{0} \leq z \leq 0.75L_{0}$, where the $x$ and $z$ directions are parallel to the solar surface, and the $y$ direction is perpendicular to it. In our 3D model, this domain is resolved by a $900\times1200\times 90$ grid. Our 2D model has the same domain size and spatial resolution in the $xy$ plane of the 3D models. Adaptive mesh refinement is not used in this pilot study. Refining and de-refining mesh can lead to the generation of artificial MHD waves, which can make it difficult to study the excitation of the ALT oscillation.
The initial density distribution is given as
\begin{eqnarray}
    \rho(x,y,z) = \rho_{\rm{chr}} + ( \rho_{\rm{cor}} -  \rho_{\rm{chr}}) \nonumber \\ \times \frac{1}{2}\left[ \tanh\left( \frac{y-h_{\rm{TR}}}{w_{\rm{TR}}}\right) +1\right],
\end{eqnarray}
where $\rho_{\rm{chr}}=10^5 \rho_{0},\ \rho_{\rm{cor}}=\rho_{0},\ h_{\rm{TR}}=1.0L_{0},\ w_{\rm{TR}}=0.2L_{0}$. The initial pressure is uniform in space, and $p(x,y,z)=p_{0}$. The initial magnetic field is assumed to be a force-free field and is described as
\begin{eqnarray}
    B_{x}(x,y,z) = 0,\\
    B_{y}(x,y,z) = B \tanh(x/w), \\
    B_{z}(x,y,z) = B/ \cosh (x/w),
\end{eqnarray}
where $B=3.92B_{0},\ w=0.5L_{0}$. The initial plasma $\beta$ is spatially uniform and is set to be $0.13$.

To induce magnetic reconnection, we adopted a spatially localized resistivity in the form of
\begin{eqnarray}
    \eta(x,y,z) = \eta_{0}\exp \left[-\left( \frac{\sqrt{x^2 + (y-h_{\eta})^2}}{w_{\eta}}\right)^2\right],
\end{eqnarray}
where $w_{\eta}=1.0L_{0},\ h_{\eta}=20L_{0},\ \eta_{0}=0.01L_{0}^2/t_{0}$. With this model, a Petschek-type magnetic reconnection with a single X-point \citep{Petschek1964NASSP} is established. Therefore, the time-variability found in our simulations is not induced by plasmoids but by other mechanisms.

\section{Results} \label{sec:results}


\subsection{Overview of the 3D model and \\
Brief Comparison with Previous 2D Models}

\begin{figure*}
    \centering
    \includegraphics[width=1.95\columnwidth]{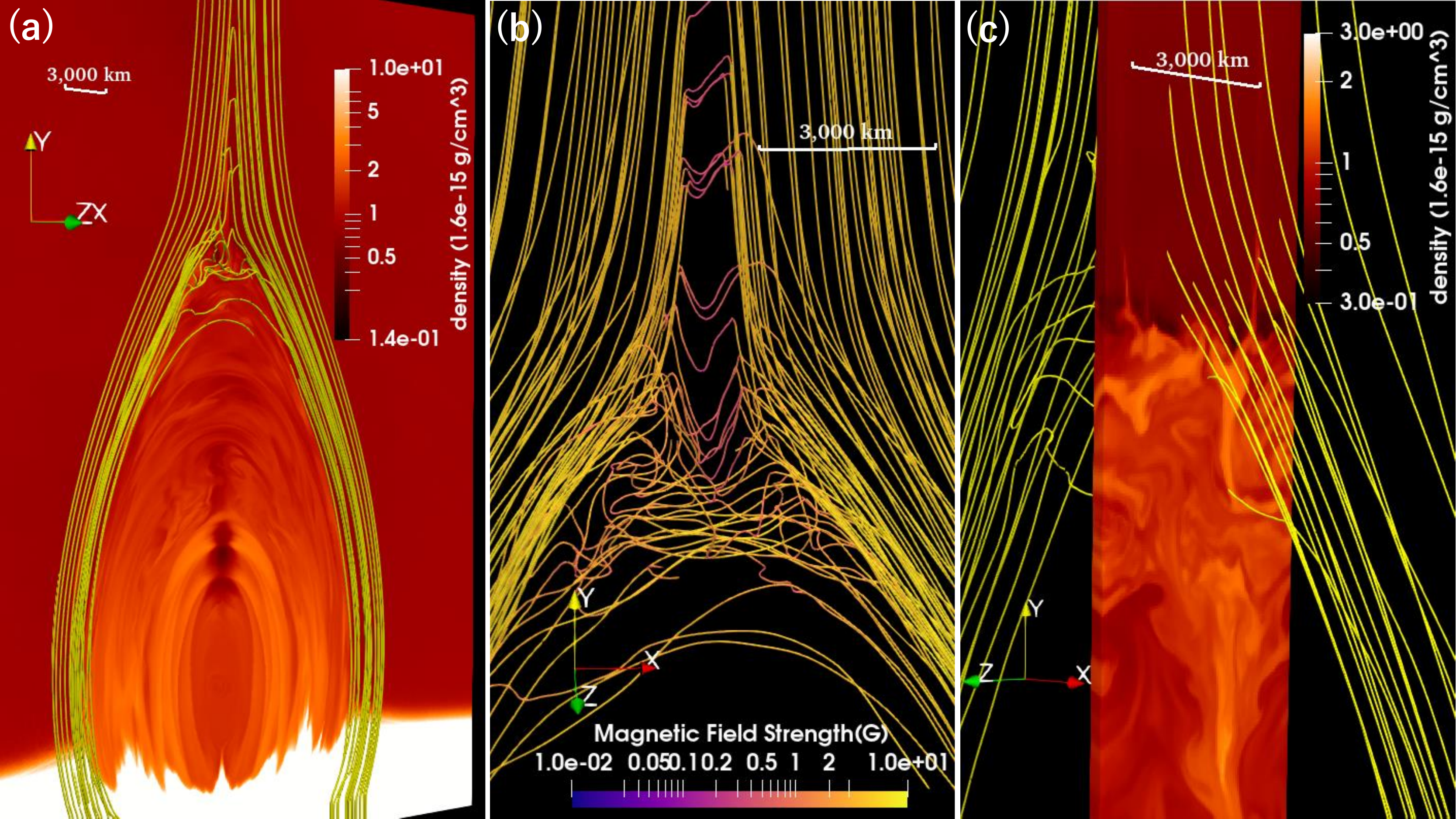}
    \caption{The 3D snapshots of a simulation at $t=660$~s. Panel (a) shows the mass density and magnetic field lines of the flare loop system. Panel (b) indicates the structure of field lines and the field strength around the ALT region. The color of the lines denotes the field strength. Panel (c) displays the density on the cross-section at $x=0.43L_{0}$ and magnetic field lines.}
    \label{figure:3d}
\end{figure*}

After the simulation begins, the flare loops start to develop around $t=280$~s. The flare loops show a bouncing motion and then form a compact ALT region. The ALT region is filled with turbulence after $t\approx 500$~s.
Figure \ref{figure:3d} presents an overview of our 3D simulation after the development of turbulence. Panel (a) shows the $xy$ slice of the density distribution and magnetic field lines (yellow lines) after the development of the flare loop system. The collimated reconnection outflow impinges on the reconnected field lines, forming a complex ALT region. Panel (b) highlights the magnetic field structure around the ALT region. Disordered fields are prominent there, suggesting the development of turbulence. Panel (c) displays the density structure in the $yz$ plane at $x=0.43L_0$, slightly shifted from the center. It is shown that the density in the ALT region is fluctuating. As we will see later, the distribution of the turbulent flows is highly inhomogeneous even though the ALT region is much smaller than the system size.

\begin{figure*}
    \centering
    \includegraphics[width=1.7\columnwidth]{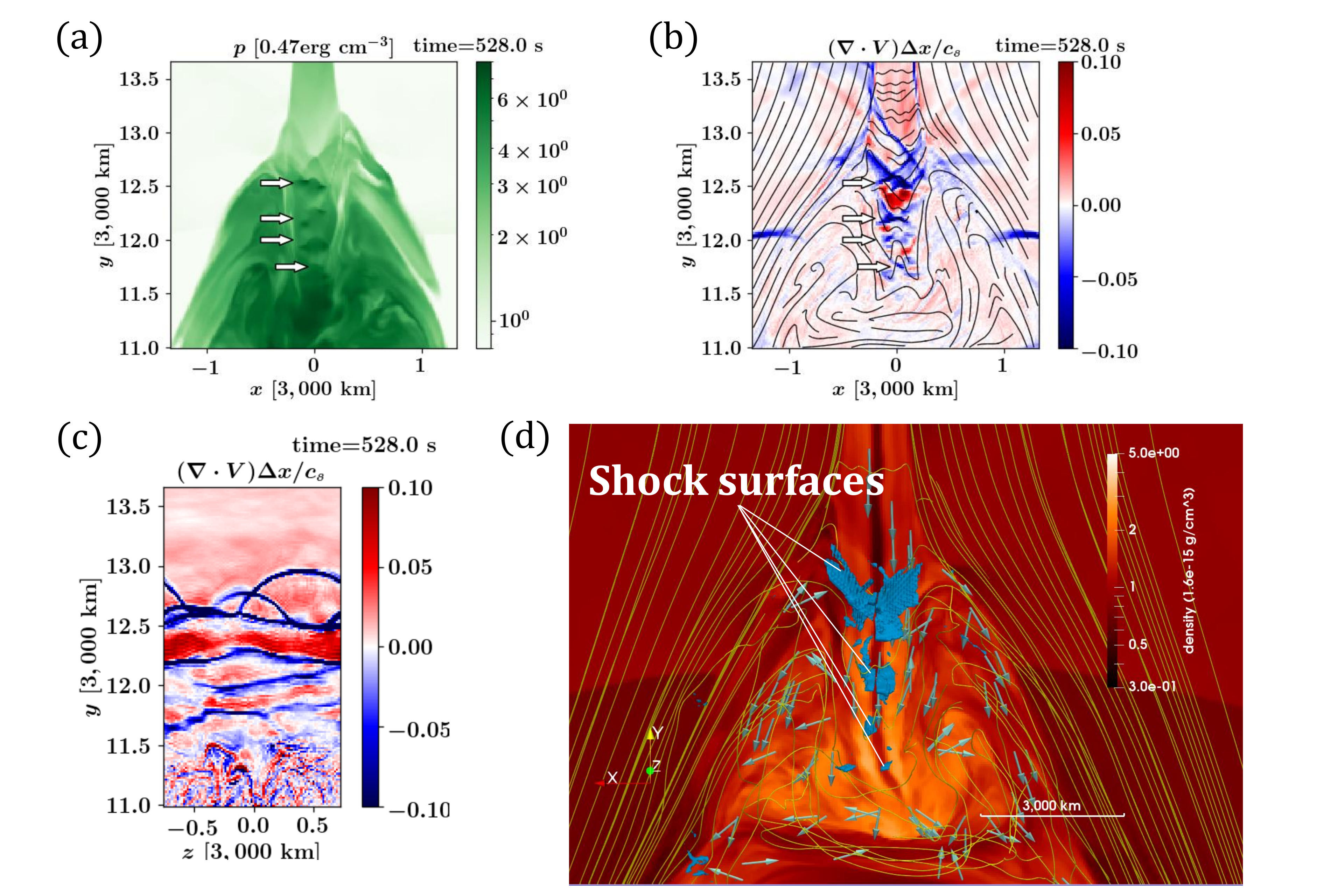}
    \caption{The shock structure around the ALT region. Panel (a) shows the pressure distribution. In panel (b), the color shows $\nabla \cdot \bm{v}$ normalized by the sound speed $c_{\rm s}$ and the mesh size $\Delta x$ in the $xy$ plane ($z=0$). Solid lines indicate the projected magnetic field structure. Arrows in Panels (a) and (b) indicate the locations of multiple termination shocks. Panel (c) is the same as (b), but in the $yz$ plane ($x=0$). Panel (d) displays the 3D image of the shock surfaces. The blue regions indicate the regions where $(\nabla \cdot \bm{v})\Delta x/c_{s} \leq -0.25$. The background color shows the mass density at $z=0$. The yellow lines denote magnetic field lines. The arrows show the direction of the velocity field in the ALT region (the size does not indicate the speed). Time is $t=528$ s for all panels.}\label{figure:shock}
\end{figure*}

Figure \ref{figure:shock} shows the termination shock structure at $t=528\ \rm{s}$. Panel (a) displays the pressure distribution at $z=0$. In panel (b), the color denotes the divergence of the velocity field normalized by the local sound speed, $(\nabla\cdot\bm{v})\Delta x /c_{\rm s}$, at $z=0$, where $c_{\rm s}$ is the adiabatic sound speed, and $\Delta x\sim 50\ \rm{km}$ is the mesh size. Shocks are highly compressed regions and are highlighted as blue linear structures. 
Arrows in panels (a) and (b) indicate the locations of multiple termination shocks. Panel (c) shows the normalized divergence of the velocity field but at the plane $x=0$. We can find that the reconnection outflow penetrates the ALT region to form multiple shocks. The formation of multiple shocks is also found in previous 2D simulations \citep{Takasao2015ApJ,Takasao2016ApJ,Zhang2022RAA}. Our result demonstrates that multiple shocks also form even in three-dimension.
Panel (d) of Figure \ref{figure:shock} represents the three-dimensional shock structure. The multiple shocks are indicated by the blue regions. Although the shock structure in the $xy$ slice is similar to that of previous 2D models, the shock surfaces are not straight in the $z$ direction at all. This shock structure is a result of the complicated flow pattern excited in the above-the-loop-top region.

\subsection{ALT oscillation}
Previous 2D models found that the backflow of the reconnection outflow can excite the ALT oscillation, even when the reconnection outflow is a quasi-steady laminar flow \citep{Takasao2016ApJ,Takahashi2017ApJ}. 
If we allow asymmetric motions about $x=0$, ALT oscillation tends to occur in an asymmetric manner because of an imbalance in the restoring force between the two arms of the magnetic tuning fork \citep{Takahashi2017ApJ}. Recent spectroscopic observations of the ALT oscillation by \citet{Reeves2020ApJ} also suggest an asymmetric oscillation; if we observe the symmetric horizontal velocity field from the side, we will not detect a significant Doppler shift.

\begin{figure}[h]
\begin{center}
    \centering
    \includegraphics[width=\columnwidth]{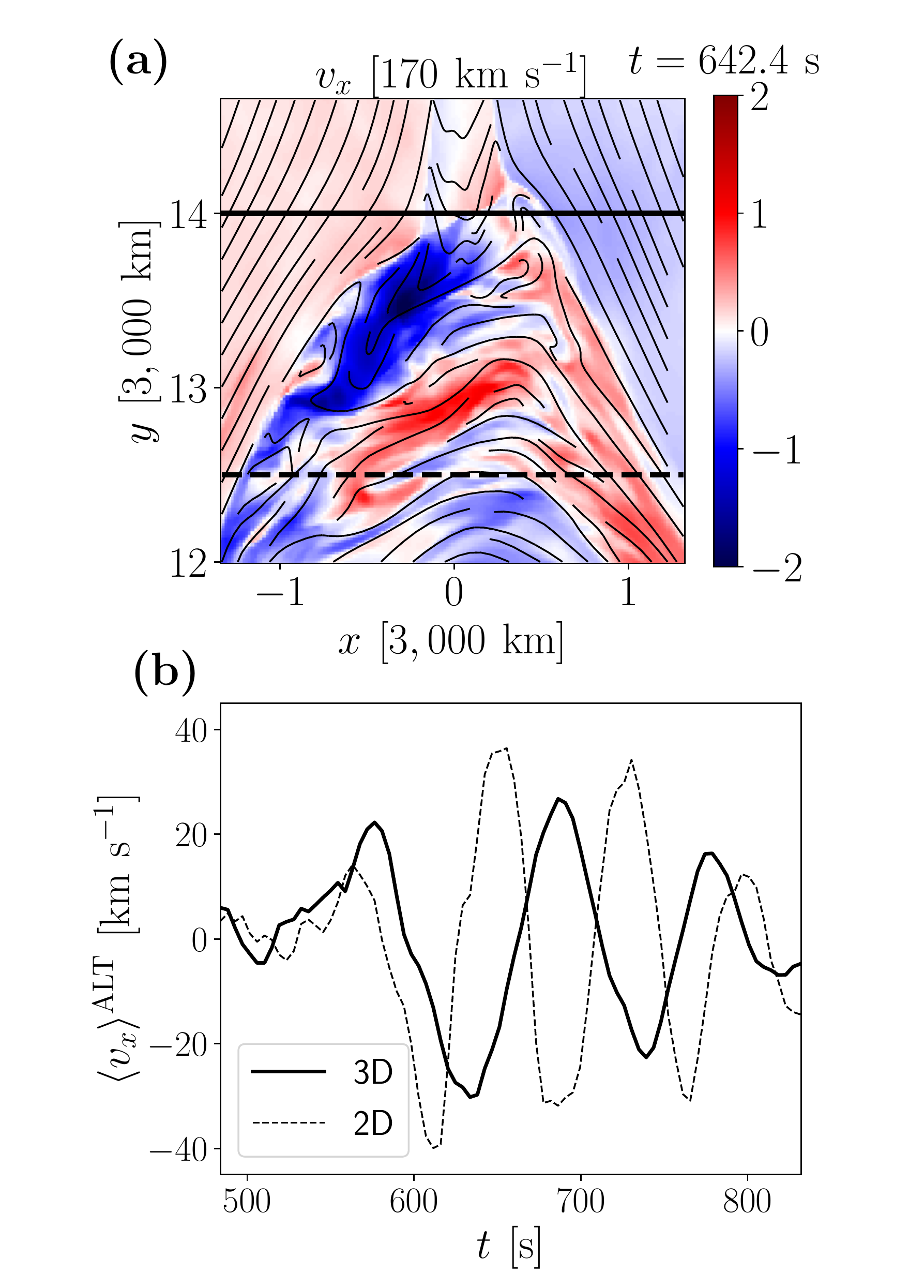}
    \caption{Panel (a) shows the spatial distribution of $v_{x}$ around the ALT region at $t=642.4$~s. The horizontal solid and dashed lines indicate $y_{\rm{top}}$ and $y_{\rm{btm}}$, respectively. Black curves denote projected magnetic field lines. Panel (b) shows the time evolution of the emission-measure-weighted horizontal velocity, $\langle v_x\rangle^{\rm ALT}(t)$. The definition is given in the main text. The solid and dashed lines display the results for 3D and 2D models, respectively. An animation of the top panel of this figure is available. The animation shows the evolution of not only $v_x$ but also $\rho$ and $\beta$ from 0 to 831.6~s. The real-time duration of the animation is 8~s.}
 \label{figure:oscillation}
\end{center}
\end{figure}

We examine the ALT oscillation in the following manner. We track the top of the ALT region and denote the height as $y_{\rm top}(t)$. We define
\begin{align}
    Dv_y&\equiv \left|\frac{dv_y}{d y}\right| \frac{\Delta y}{c_{\rm s}}\\
    Dp&\equiv \left|\frac{dp}{dy}\right|\frac{\Delta y}{p},
\end{align}
where $\Delta y$ is the cell size in the $y$ direction.
$Dv_y$ and $Dp$ measure the jumps in the reconnection jet velocity and the gas pressure, respectively. Considering that MHD fast-mode shocks are formed at the top of the ALT region, $y_{\rm top}(t)$ is defined as the maximum height at which $Dv_y\cdot Dp>D_c^2$, where $D_c$ is a nondimensional number. The magnitude of $D_c$ represents a rough threshold for the sizes of the jumps in $v_y$ and $p$ across a cell of the fast-mode shocks. After some trials, we find that the value of $D_c= 0.1$ works well for tracking.
We define the ALT region as the region where the reconnection jet penetrates into the post-flare loops. With this definition, the typical vertical size of the ALT region is approximately $1L_0{\rm -}1.5L_0$. Considering this, we set the bottom height of the ALT region to be $y_{\rm btm}(t)=y_{\rm top}(t)-1.5L_0$.
The left panel of Figure~\ref{figure:oscillation} indicates the locations of $y_{\rm top}(t)$ and $y_{\rm btm}(t)$ as the solid and dashed lines, respectively. The panel displays the $v_x$ distribution, where we can discern how the penetrating jet is refracted. We provide an animation demonstrating that we can successfully track the ALT region with this method.

Next, we calculate the emission-measure-weighted horizontal velocity ($v_x$) in the ALT region, which will be a good indicator of the Doppler velocity in spectroscopic observations. We first average the velocity in the $x$ and $z$ directions:
\begin{align}
    \langle v_{x}\rangle^{z}(t,x,y) \equiv \frac{1}{L_{z}}\int_{-L_z/2}^{L_z/2} v_{x}(t,x,y,z)dz,\\
    \langle \rho\rangle^{z}(t,x,y) \equiv \frac{1}{L_{z}}\int_{-L_z/2}^{L_z/2} \rho(t,x,y,z)dz,\\
    \langle v_{x}\rangle^{xz}(t,y)
    \equiv \frac{\int_{x_{\rm ALT,L}}^{x_{\rm ALT,R}} \langle \rho\rangle^{z}(t,x,y)^2 \langle v_{x}\rangle^{z}(t,x,y)  dx}{\int_{x_{\rm ALT,L}}^{x_{\rm ALT,R}} \langle \rho\rangle^{z}(t,x,y)^2 dx },
\end{align}
where, $L_{z}=1.5L_{0}$ (the simulation domain size in the $z$ direction), $x_{\rm ALT,L}=-2.0L_{0}$, and $x_{\rm ALT,R}=2.0L_{0}$, respectively. To obtain the averaged value for the ALT region, $\langle v_{x}\rangle^{xz}(t,y)$ is averaged in the range of $y_{\rm btm}(t)\le y \le y_{\rm top}(t)$:
\begin{align}
    \langle v_{x}\rangle^{\rm ALT}(t)=\frac{1}{y_{\rm top}(t)-y_{\rm btm}(t)}\int_{y_{\rm btm}}^{y_{\rm top}}\langle v_{x}\rangle^{xz}(t,y)dy.
\end{align}
Panel (b) of Figure~\ref{figure:oscillation} displays the result (solid line).
Approximately two cycles of oscillation are found. The period is $\sim 100$~s. The dashed line is for the 2D model. The 3D model shows a similar but longer oscillation period than that of the 2D model. Considering that the ALT oscillation is driven by the horizontal backflow, the slightly smaller velocity amplitude in the 3D model probably results in a longer period. Turbulence in the 3D model seems to reduce the coherent velocity. However, our 3D model demonstrates that the ALT oscillation indeed occurs in three-dimension.

\subsection{Local generation of turbulence in the ALT region} \label{sec:instability}
\begin{figure*}
\begin{center}
    \centering
    \includegraphics[width=2\columnwidth]{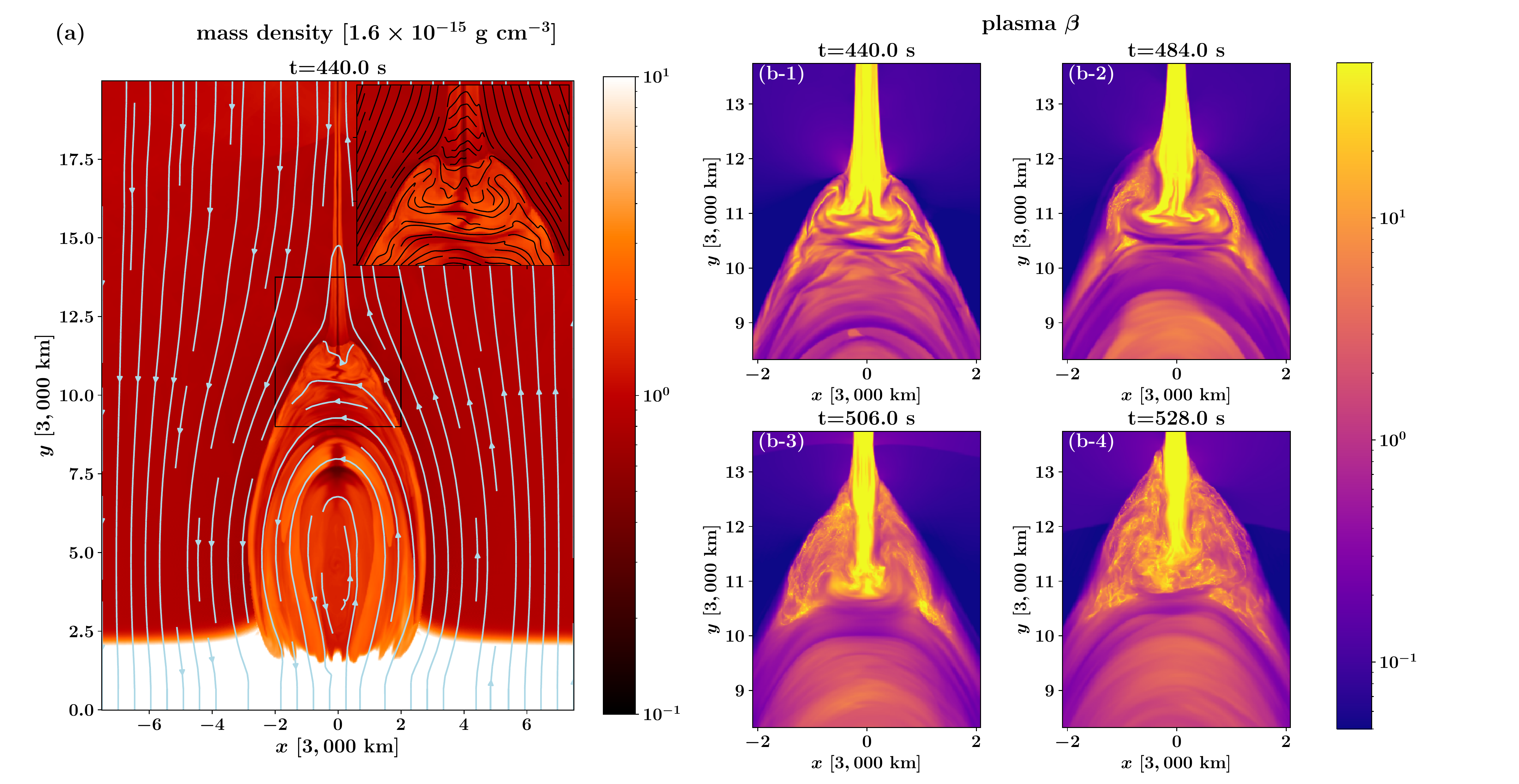}
    \caption{Development of turbulent flows around the ALT region. Panel (a): the density with the projected field lines in the $xy$ plane ($z=0$). 
    Panels (b-1) to (b-4): the plasma $\beta$ distributions around the ALT region at different times. The box size is indicated by the black square in Panel (a). The center of the box is shifted with time to cover the ALT region. }\label{figure:turbulence_beta}
\end{center}
\end{figure*}

Figure \ref{figure:turbulence_beta} demonstrates the local generation of turbulence in the ALT region.
Panel (a) shows the density distribution with the projected field lines in the $xy$ plane ($z=0$). The inset shows the enlarged image of the ALT region.
Panels (b-1) to (b-4) show the time evolution of the plasma $\beta$ around the ALT region. As time proceeds, fine-scale structures develop around the two arms of the magnetic tuning fork (see Panels (b-2) and (b-3)). Eventually, the ALT region is filled with turbulent flows (Panel (b-4)).
The reconnection outflow itself is a laminar flow and contains no plasmoids. Therefore, the turbulence should be locally excited in the ALT region.

The ALT region seems to be unstable to both pressure-driven (or bad-curvature-driven) instabilities and a centrifugally driven Rayleigh-Taylor-type (RT) instability. Both are related to the particular magnetic geometry of the ALT region. In the following, we describe where these mechanisms operate.

\begin{figure*}
    \centering
    \includegraphics[width=18.0cm]{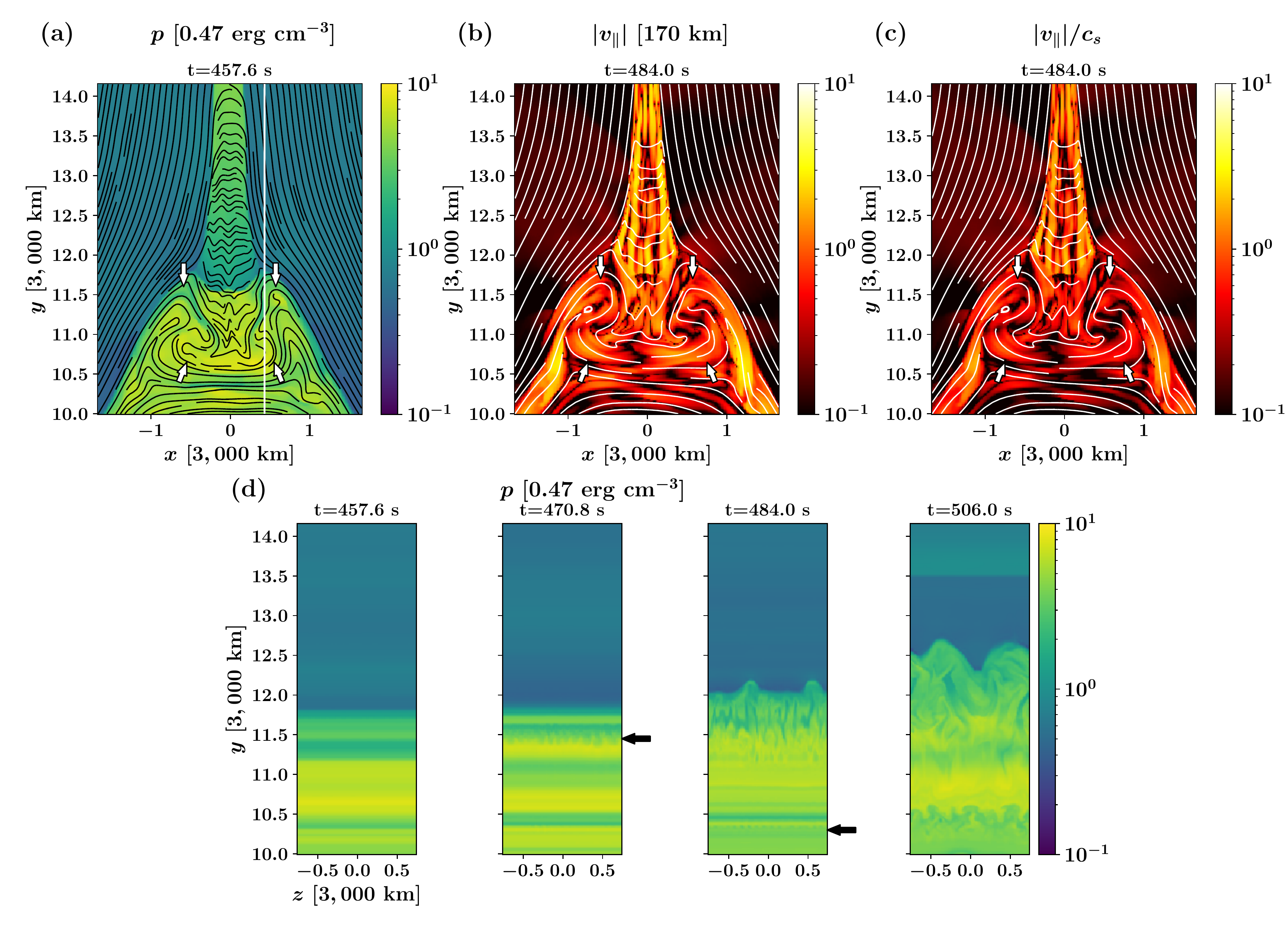}
    \caption{Panel (a) shows the pressure distribution around the ALT region, where four bad-curvature regions are indicated by white arrows. Black lines represent the magnetic field lines projected onto this plane. Panel (b) displays the magnitude of the velocity along a magnetic field line, $v_{\parallel}=|\bm{v} \cdot \bm{B}|/|\bm{B}|$. Panel (c) indicates $v_{\parallel}$ normalized by the sound speed $c_{s}$. Panel (d) displays the pressure distributions at four different times in the $yz$ plane indicated by the white line in Panel (a), where the development of instabilities around bad-curvature regions is shown. The heights are indicated by black arrows.}\label{figure:inst}
\end{figure*}

The ALT region contains a high-pressure plasma, as the kinetic energy of the reconnection outflow is converted into heat (see Panel (a) of Figure~\ref{figure:shock}). The high-pressure plasma is confined by a curved magnetic field (see the regions indicated by white arrows in Figure~\ref{figure:inst}), and such a plasma configuration can be unstable to pressure-driven (or bad-curvature-driven) instabilities. 
Bad curvature is defined by the relation between the magnetic field curvature vector and the pressure gradient vector. The magnetic field curvature vector $\bm{\kappa}$ is defined as 
\begin{equation}
    \bm{\kappa} = (\bm{b}\cdot\nabla)\bm{b},
\end{equation}
where $\bm{b}=\bm{B}/|\bm{B}|$. Plasma has bad curvature when
\begin{equation}
    \bm{\kappa}\cdot \bm{\nabla}p>0
\end{equation}
and can become unstable \citep[e.g.][]{Freidberg2014}. Both interchange and undular modes can grow, and if we can ignore magnetic shear, modes with a larger $k_\perp$ will grow more rapidly in both cases. The instabilities are essentially driven by the gas pressure gradient force. An important example of the interchange or flute modes is the pressure-driven version of the RT instability.
The growth rate of pressure-driven modes, $\gamma_{\rm grow,p}$, is written approximately as
\begin{equation}
    \gamma_{\rm grow,p}\sim \frac{c_{\rm s, ALT}}{\sqrt{L_p R_{\rm c}}},\label{eq:growth_rate1}
\end{equation}
where $L_p^{-1} \equiv |\nabla p|/p$ is the pressure-gradient length scale and $R_{\rm c}$ denotes the curvature of the magnetic field. $c_{\rm s,ALT}$ denotes the sound speed in the ALT region. Therefore, the growth rate is larger when the high-pressure gas is confined by a more highly curved magnetic field with a larger pressure gradient.

The above instability mechanisms ignore the effect of the plasma flows in the unperturbed state. However, we find transonic plasma flows along a curved magnetic field, which can induce a centrifugally driven RT instability. Figure~\ref{figure:inst} shows the velocity component parallel to a magnetic field, $\bm{v}\cdot\bm{B}/B$. The figure indicates that backflowing plasma is flowing with a transonic speed along a curved magnetic field (see the regions indicated by red arrows in panel (a)). Because of curvature, the plasma feels centrifugal force. Therefore, if there is a density contrast in the flows, we expect a centrifugally driven Rayleigh Taylor instability. The growth rate of this instability, $\gamma_{\rm grow,c}$, is estimated to be
\begin{equation}
    \gamma_{\rm grow,c}\sim \sqrt{\frac{g_{\rm eff}}{R_c}}\approx \frac{v_{\rm para}}{R_{\rm c}},\label{eq:growth_rate2}
\end{equation}
where $g_{\rm eff}=v_{\rm para}^2/R_c$ is the effective acceleration due to centrifugal force. We expect the growth of interchange modes with high wavenumber in $z$ direction.

In summary, the ALT region seems to be unstable to two types of instabilities (pressure-driven instabilities and the centrifugally driven RT instability), and the growth rate of the most rapidly growing mode will be 
\begin{equation}
    \gamma_{\rm grow} \sim \max{(\gamma_{\rm grow,p},\gamma_{\rm c})}=\max{(\frac{c_{\rm s,ALT}}{\sqrt{L_p R_{\rm c}}}, \frac{v_{\rm para}}{R_{\rm c}})}\label{eq:max_growth_rate}
\end{equation}
If we can assume $v_{\rm para}\approx c_{\rm s,ALT}$ and $L_p\approx R_{\rm c}$ in the parameter space of interest, then 
\begin{eqnarray}
    \gamma_{\rm grow}\sim c_{\rm s,ALT}/R_{\rm c}.\label{eq:max_growth_rate_approx}
\end{eqnarray}
As the backflow is driven by the gas pressure in the ALT region, $v_{\rm para}\approx c_{\rm s,ALT}$ will be a reasonable assumption. $L_p$ can be significantly smaller than $R_{\rm c}$ because of the sharp boundary inside and outside the ALT region. Therefore, the actual growth rate can be larger than the estimate of Equation~(\ref{eq:max_growth_rate_approx}).

\begin{figure}
    \centering
    \includegraphics[width=\columnwidth]{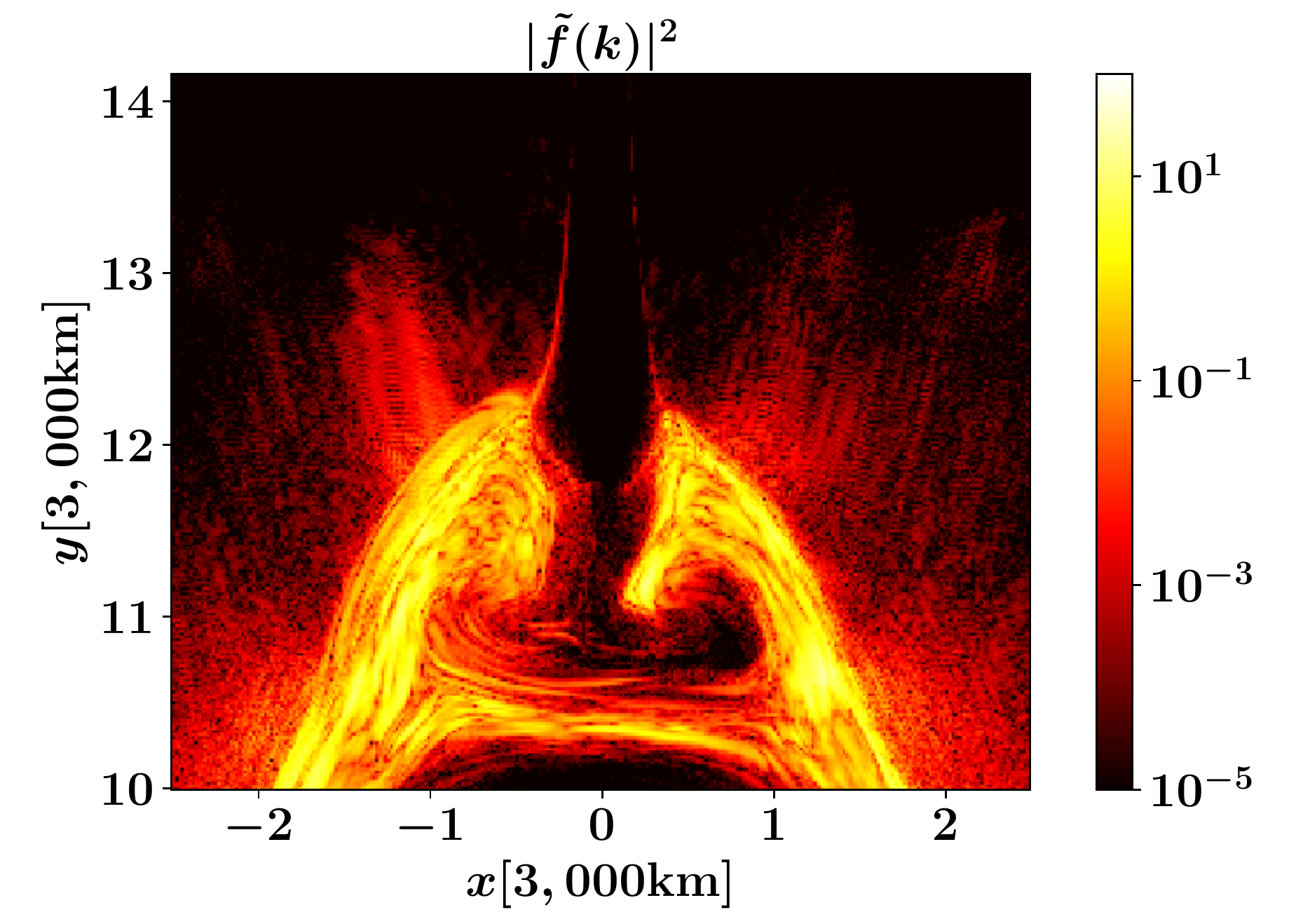}
    \caption{The spatial distribution of the Fourier power of the density fluctuation. The power corresponding to a wave number of $k=1.12\ [\rm{cell\ size^{-1}}]$ is shown. The data is taken at $t=484\ \rm{s}$.}
 \label{figure:power_noHC}
\end{figure}

Figure \ref{figure:inst} demonstrates the growth of the instabilities. Panel (a) shows the $xy$ cutout of the pressure distribution. Regions with bad-curvature are indicated by white arrows; the bottom end of the reconnection outflow and two arms of the magnetic tuning fork. 
Panels (b-1) to (b-4) indicate the time evolution of the pressure in the cutout plane indicated by the white line in Panel (a). The fluctuations indeed grow in the bad-curvature regions.

For a quantitative analysis, we performed the Fourier analysis of the density fluctuation in the $z$ direction. Figure \ref{figure:power_noHC} shows the spatial distribution of the Fourier power corresponding to a wave number of $k=1.12\ [\rm{cell\ size^{-1}}]$. It is difficult to identify the types of modes. Nevertheless, it is clear that the grid-scale modes have large powers in the arms of the magnetic tuning fork. The fastest growing mode has a wavelength of $\sim 6 \Delta z$. The fact that the fastest growing mode occurs at such a grid scale and in the arms is consistent with the pictures of the instabilities we consider.
The power at the bottom of the ALT region is much weaker, which demonstrates a slower growth of the instabilities there.

The development of instabilities around the bottom end of the reconnection outflow is also analyzed in previous studies \citep{Guo2014ApJ,Shen2018ApJ,Shen2022NatAs}. However, the detailed structures of the magnetic tuning fork have been overlooked. The effect of the centrifugal force due to plasma flows was also ignored in previous studies.
This study found that the instabilities in the arms of the magnetic tuning fork grow more rapidly than those around the bottom edge of the outflow. The turbulent flows develop first in the arms of the magnetic tuning fork, as shown in Figure~\ref{figure:turbulence_beta}. The turbulent regions extend in size and eventually surround the termination shocks. Therefore, the instabilities in the arms of the magnetic tuning fork have a more significant impact on the plasma structure just around the termination shocks.

\subsection{Spatial Distribution of Turbulent Flows}
We quantify the strength of turbulence by measuring the amplitudes of the fluctuations in the velocity and magnetic fields. The velocity fluctuation $\delta v$ and the magnetic field fluctuation $\delta B$ are respectively defined as
\begin{eqnarray}
    \delta v(t,x,y,z)^2 = (v_{x}(t,x,y,z)-\langle v_{x}\rangle(t,x,y))^2 \nonumber \\
    + (v_{y}(t,x,y,z)-\langle v_{y}\rangle(t,x,y))^2 \nonumber\\
    + (v_{z}(t,x,y,z)-\langle v_{z}\rangle(t,x,y))^2 \label{15}
\end{eqnarray}
and 
\begin{eqnarray}
    \delta B^2 = (B_{x}-\langle B_{x}\rangle)^2 + (B_{y}-\langle B_{y}\rangle)^2 \nonumber\\
    + (B_{z}-\langle B_{z}\rangle)^2 \label{16},
\end{eqnarray}
where $\langle a \rangle$ denotes the average value in the $z$ direction for a physical quantity $a$:
\begin{eqnarray}
    \langle a \rangle(t,x,y) = \frac{1}{L_{z}}\int_{-L_z/2}^{L_z/2} a(t,x,y,z)dz.
\end{eqnarray}

\begin{figure*}
    \centering
    \includegraphics[width=1.9\columnwidth]{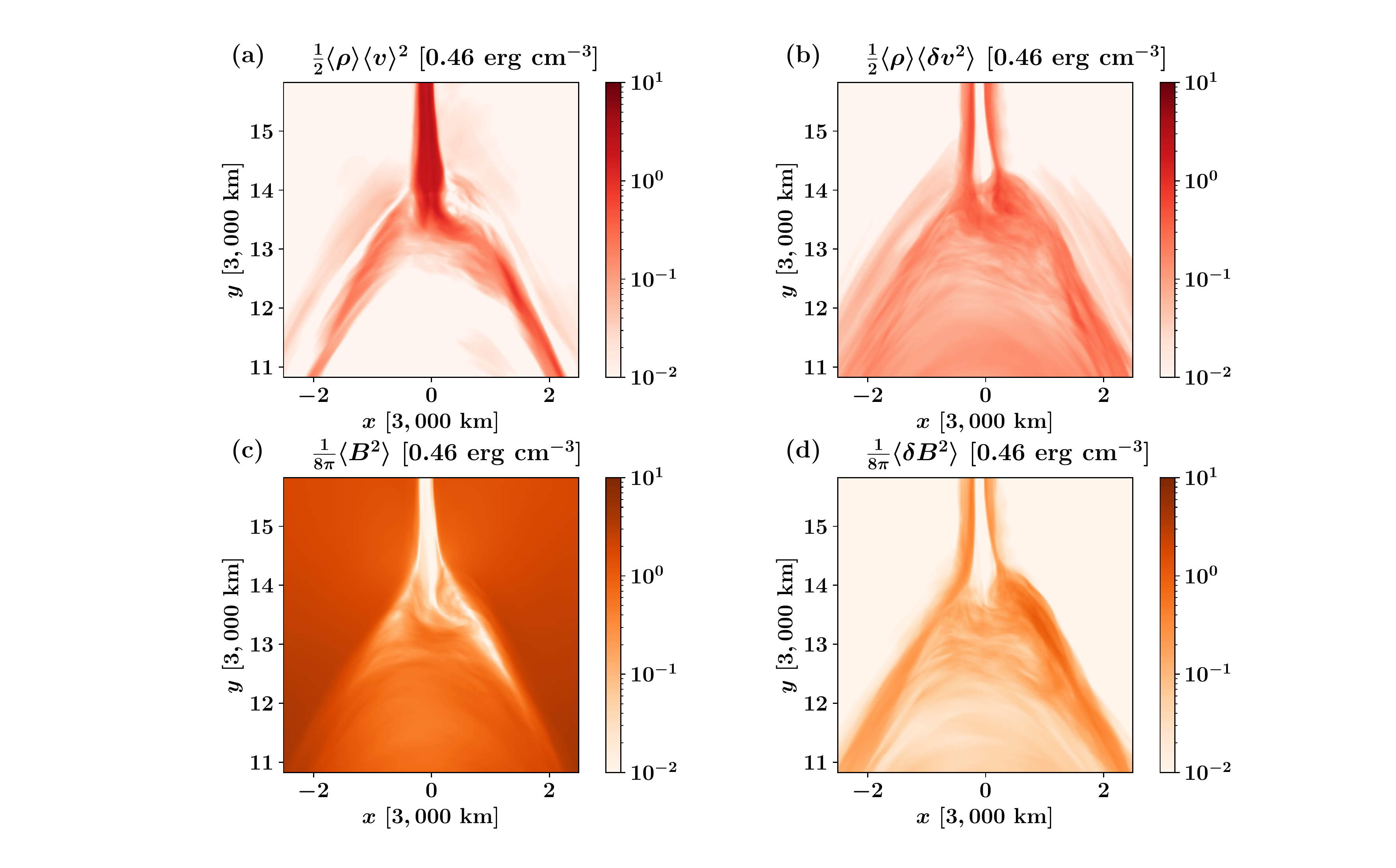}
    \caption{The kinetic and magnetic energy density distributions at $t=704\  \rm{s}$. Panels (a) and (b) show the kinetic energy density for coherent and turbulent (fluctuating) flows, respectively. Panels (c) and (d) indicate the magnetic energy densities for coherent and turbulent magnetic fields, respectively.}
    \label{figure:turblent_energy_density}
\end{figure*}

The spatial distributions of the turbulent kinetic and magnetic energy densities are shown in Figure~\ref{figure:turblent_energy_density}, respectively. The figure indicates that the turbulent energy densities take larger values in the arms of the magnetic tuning fork.

Figure \ref{figure:turblent_energy_density} displays the spatial distributions of the kinetic (top) and magnetic (bottom) energy densities. The left and right panels exhibit the coherent and turbulent components, respectively.
It is shown that the relative magnitude of the turbulent components to the coherent components is highly inhomogeneous both in the kinetic and magnetic energy densities.
The turbulent components of the kinetic and magnetic energy densities are much larger than the coherent components in the two arms of the magnetic tuning fork, which results from the local generation of turbulence via the interchange instabilities. 
The part of the reconnection outflow inside the ALT region shows a complicated structure (Figure \ref{figure:shock}), but the turbulent components are much smaller than the coherent components.

We found that the ALT oscillation promotes the local generation of turbulence in the arms of the magnetic tuning fork. The ALT oscillation is essentially a compressible process driven by the backflow of the reconnection outflow \citep{Takasao2016ApJ}. The backflow compresses the gas and magnetic field around the arms, increasing the pressure gradient. As a result, the growth rate of the interchange mode increases (Equation (\ref{eq:growth_rate1})). 
Figure \ref{figure:turblent_energy_density} shows the snapshots when the asymmetric ALT oscillation induces strong turbulence in the right arm. Our simulation demonstrates that the amplitude and the spatial distribution of turbulence vary with time because of the asymmetric ALT oscillation.

\begin{figure}
    \centering
    \includegraphics[width=\columnwidth]{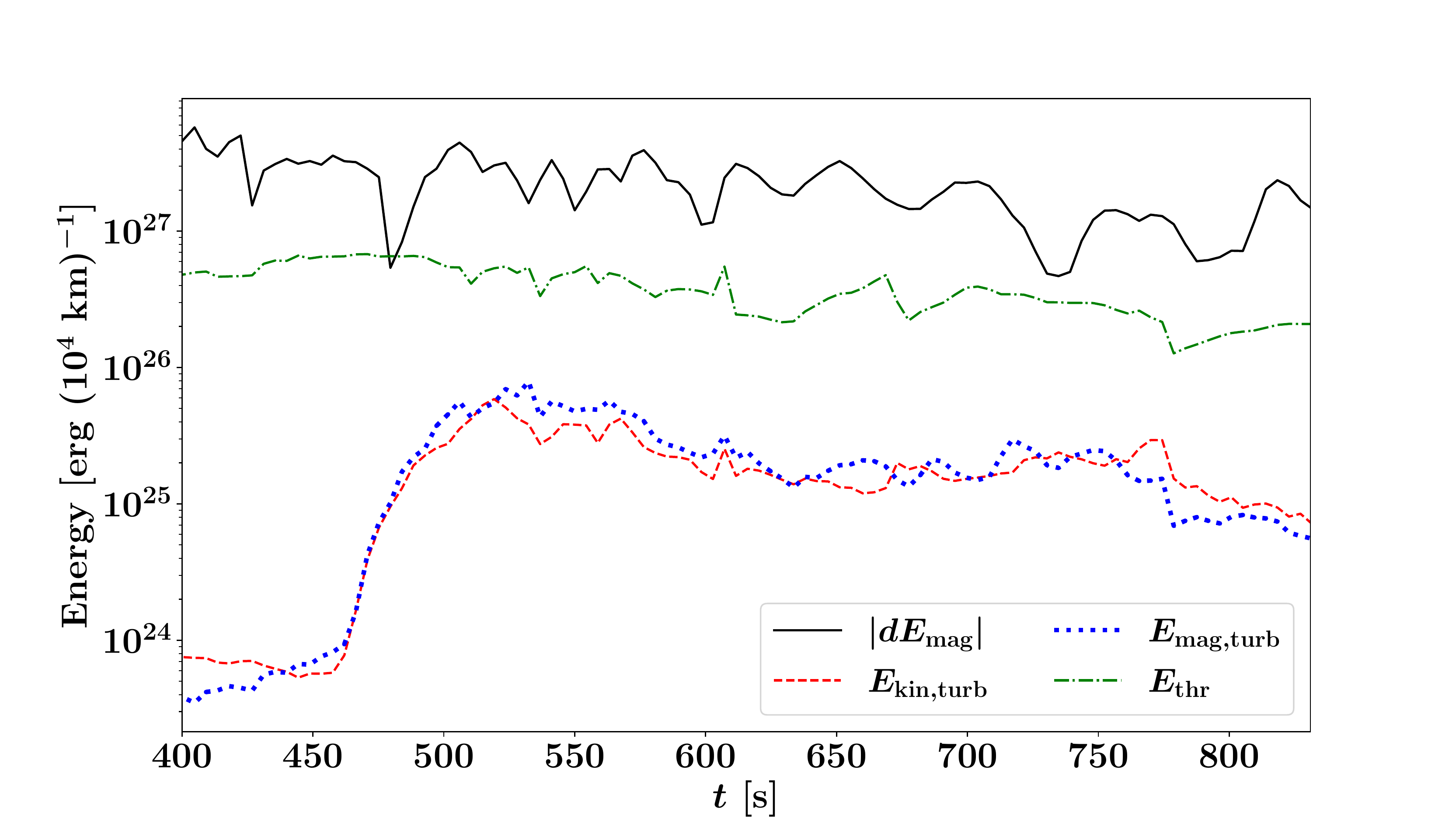}
    \caption{Time evolution of the different energies associated with turbulence in the ALT region. $|dE_{\rm mag}|$ denotes the magnetic energy released by magnetic reconnection during the Alfv\'en transit timescale. $E_{\rm kin,turb}$ and $E_{\rm mag,turb}$ are the turbulent kinetic and magnetic energies  in the ALT region, respectively. $E_{\rm thr}$ is the thermal energy in the ALT region. Note that the unit of these quantities is the energy per unit length in the $z$ direction. See the main text for more details about the definitions.}
    \label{figure:ALTenergy}
\end{figure}


\begin{figure*}
    \centering
    \includegraphics[width=19.0cm]{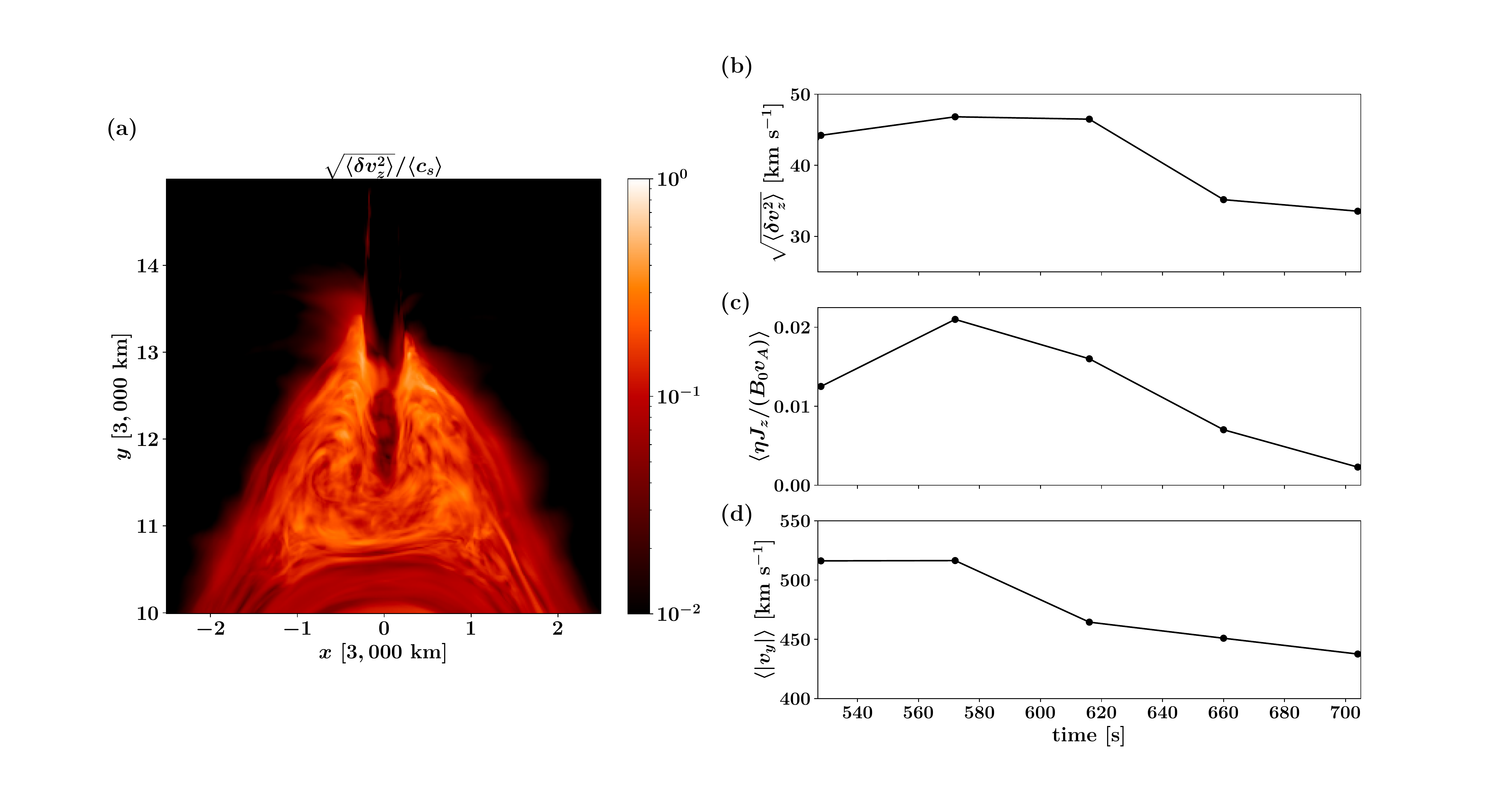}
    \caption{Panel (a) shows the spatial distribution of the acoustic Mach number of $\langle \delta v_z^2\rangle$ at $t=528$~s. Panels (b-d) show the time evolution of $\sqrt{\langle \delta v_z^2\rangle}$, the non-dimensional reconnection rate $\langle \eta J_{z}/(B_{0}v_{A}) \rangle$ at $(x,y)=(0,20L_{0})$, and the speed of the reconnection outflow $\langle |v_{y}| \rangle$ at $(x,y)=(0,y_{\rm top} + L_{0})$, respectively. These three panels share the horizontal axis.}
    \label{figure:time_vz}
\end{figure*}

We study the energy conversion from the magnetic energy released by reconnection into the turbulent kinetic and magnetic energies.
We define the quantity associated with the magnetic energy released by magnetic reconnection, $dE_{\rm mag}(t)$, as 
\begin{align}
    dE_{\rm mag}(t) = \frac{d E_{\rm mag,all}}{d t}\frac{h_{\eta}}{v_{\rm A,0}}\frac{1}{L_{z}},
\end{align}
where $E_{\rm mag,all}(t)$ is the total magnetic energy in the numerical domain and $v_{A,0}\approx 670~\rm{km~s^{-1}}$ is the Alfv\'en speed of the initial corona. $h_{\eta}/v_{A,0}(\approx 90~{\rm s})$ denotes the Alfv\'en transit timescale in the $y$ direction (a typical timescale for the reconnection outflow). 
Therefore, $dE_{\rm mag}(t)*L_z$ expresses the magnetic energy released by reconnection during the Alfv\'en transit timescale.
$dE_{\rm mag}(t)$ defines the total energy per unit length available for driving turbulence.
The turbulent kinetic and magnetic energies per unit length in the ALT region are respectively defined as
\begin{align}
    E_{\rm kin,turb}(t) =
    \iint_{\rm ALT}\frac{1}{2} \langle \rho \rangle \langle \delta v^2 \rangle (t,x,y) dxdy,
\end{align}
and
\begin{align}
    E_{\rm mag,turb}(t) =
    \iint_{\rm ALT} \frac{1}{8\pi} \langle  \delta B^2 \rangle(t,x,y) dxdy.
\end{align}
Where, the domain of integration is the region satisfying $y_{\rm btm}(t) \leq y \leq y_{\rm top}(t)$ and $\langle p \rangle(t,x,y)\geq p_{0}$. We also define the thermal energy $E_{\rm thr}$ per unit length in the ALT region as
\begin{align}
    E_{\rm thr}(t) =
    \iint_{\rm ALT} \frac{1}{\gamma-1}\langle  p \rangle(t,x,y) dxdy.
\end{align}
The thermal energy gives the upper limit of the energy of turbulence produced by the pressure-driven instabilities.

Figure \ref{figure:ALTenergy} shows the evolution of the energies defined above. Comparing $|dE_{\rm mag}|$ and $E_{\rm thr}$, one will find that approximately a few 10\% of the magnetic energy released by reconnection is converted into the heat in the ALT region. We note that the thermal energy will be smaller if heat conduction cooling is activated.
$E_{\rm kin,turb}$ and $E_{\rm mag,turb}$ are comparable to each other and much smaller than $|dE_{\rm mag}|$. They are approximately a few percent of $|dE_{\rm mag}|$.
This conversion efficiency is comparable to the observational estimation in \citet{Kontar2017PhRvL}, where they estimate it to be $\sim(0.5-1)$\% (although their estimation of the released magnetic energy is an order-of-magnitude estimation based on the limited observational information).

We investigate the time evolution of the turbulent velocity in the ALT region. As an indicator of the turbulent velocity, we examine $\sqrt{\langle \delta v_z^2 \rangle}$. Figure \ref{figure:time_vz} (a) displays the Mach number of the turbulence, which is defined as the ratio of $\sqrt{\langle \delta v_z^2 \rangle}$ to the averaged local sound speed, $\langle c_{\rm s}\rangle$. The Mach number approximately ranges 0.1-0.3 in the ALT region, except for the part of the reconnection outflow within the ALT region. To study the time evolution of the turbulent velocity, we took the spatial average of the turbulent velocity within the ALT region (in the region where $y_{\rm btm}(t)\le y \le y_{\rm top}(t)$ and $\langle p \rangle \geq p_{0}$). The result is shown in Panel (b), where one can find a reduction in the turbulent velocity from $\sim 48$~km~s$^{-1}$ to $\sim 32$~km~s$^{-1}$.

Considering that the turbulence is produced by the injection of the reconnection jet into the ALT region, we examine the relation between the reconnection and the turbulence.
Panel (c) of Figure \ref{figure:time_vz} displays the nondimensional reconnection rate averaged in the $z$ direction, $\langle \eta J_{z}/(B_{0}v_{A}) \rangle$ at $(x,y)=(0,20L_{0})$, where $J_{z}$ is the $z$ component of the electric current density vector. The reconnection rate monotonically decreases after $t\approx 570$~s. The panel (d) displays the speed of the reconnection outflow averaged in the $z$ direction. It is measured at $(x,y)=(0,y_{\rm top}(t)+1L_0)$.
The panel (d) shows that the jet speed decreases as the reconnection rate does. The smaller jet speed leads to a smaller pressure and backflow speed in the ALT region. Considering that the turbulence is driven by both pressure and centrifugal force, the decrease in the reconnection jet speed results in the reduction in the turbulent velocity.
The termination shock structure may have some effect on the reduction of the turbulence speed.
\citet{Takasao2016ApJ} pointed out that the backflow speed is smaller after the shock is nearly a horizontal shock.
The uppermost termination shock in our simulation is nearly a horizontal shock after $t\approx 690$~s.

\begin{figure}
    \centering
    \includegraphics[width=\columnwidth]{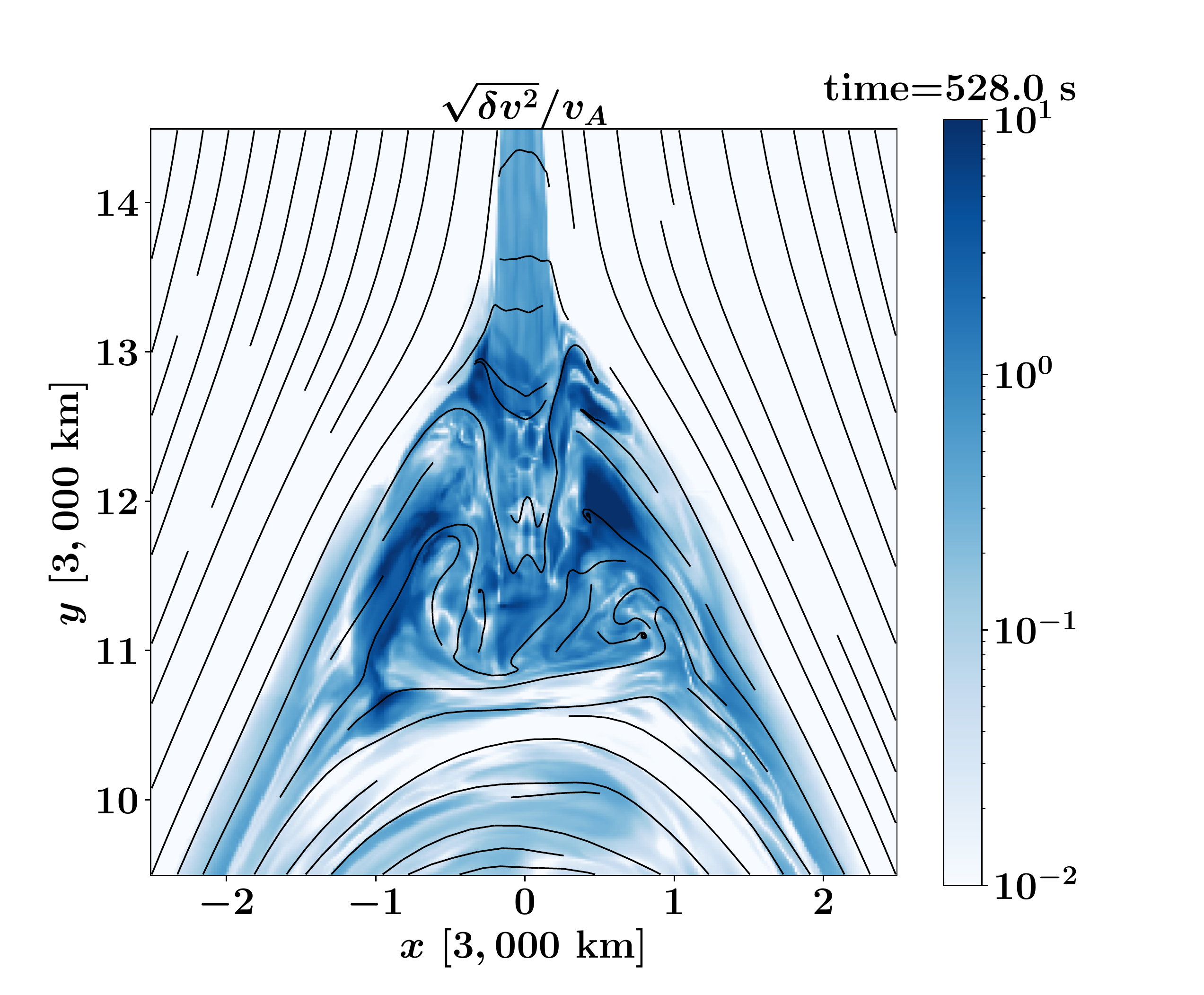}
    \caption{The distribution of the Alfv\'en Mach number of the turbulent velocity at $t=528\ \rm s$. The black lines denote projected magnetic field lines.}
    \label{figure:Alfven_Mach}
\end{figure}

The Alfv\'en Mach number of the fluctuation is also studied. Figure~\ref{figure:Alfven_Mach} displays $\sqrt{\delta v^2}/\langle v_{\it A}\rangle$. The value exceeds unity in the ALT region, indicating the development of super Alfv\'enic turbulence. In addition, fluctuations propagate from the ALT region to the foot-points of the flare loops in the form of large-amplitude MHD waves. \citet{Kigure2010PASJ} investigated the energy transport by MHD waves produced around the reconnection region. This study demonstrates that the released magnetic energy is carried away by MHD waves not only from the reconnection region but also from the turbulent ALT region. The impact of the energy transport will be investigated in the future.

\subsection{Possible impact of turbulence \\
on the magnetic mirror trap}

\begin{figure*}
    \centering
    \includegraphics[width=18.0cm]{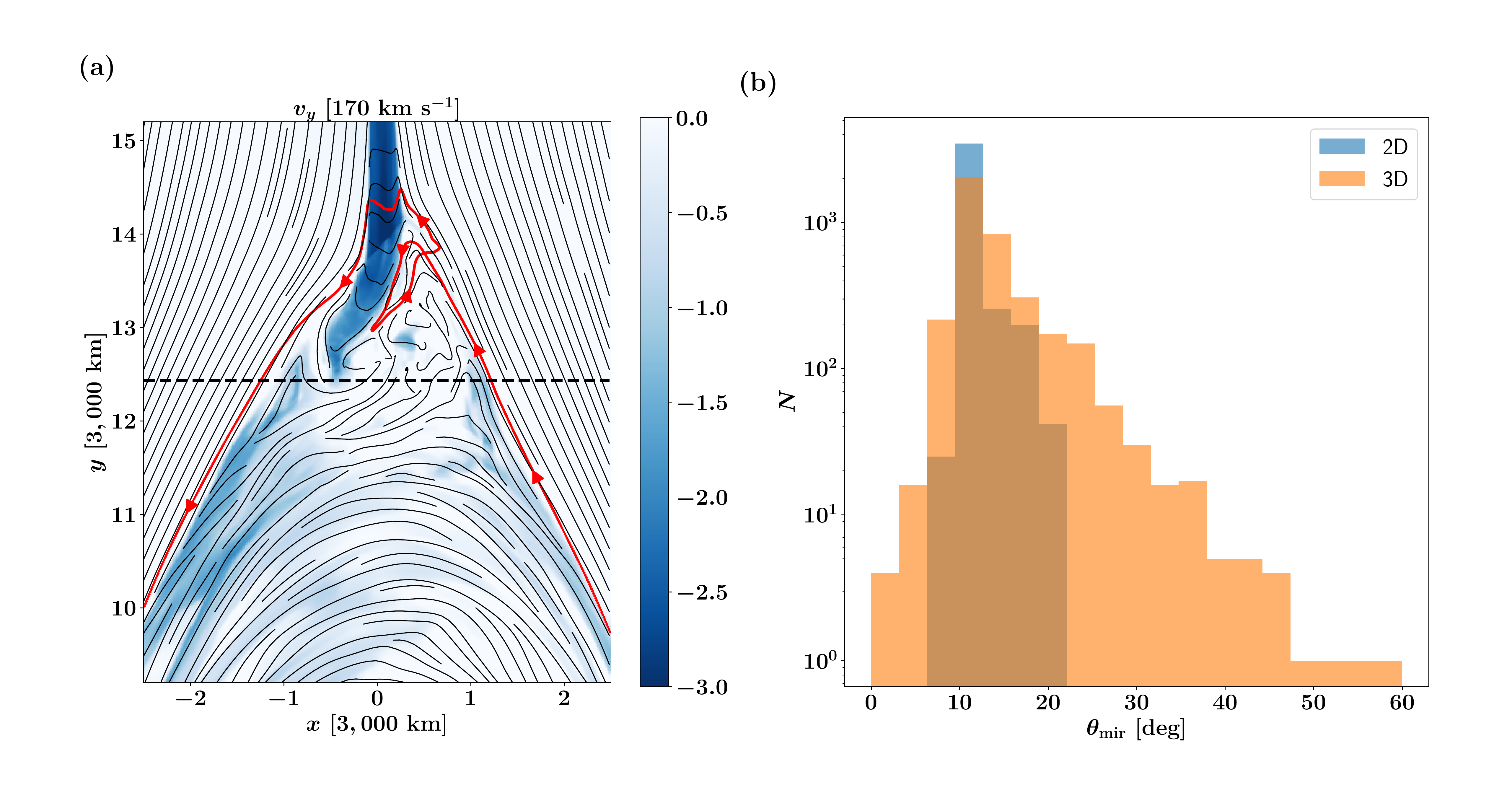}
    \caption{Panel (a) shows the spatial distribution of $v_{y}$(color map), magnetic field lines (black lines), and an example of tracked magnetic field lines (the red line) at $t=616$ s. The arrows along the red line denote the direction of the magnetic field. The dashed line indicates the line $y=y_{\rm{btm}}$. Panel (b) displays the histogram of the minimum pitch angle $\theta_{\rm{mir}}$ for magnetic mirror reflection at $t=616$ s. The blue and orange bars show the results of the 2D and 3D models, respectively.}
    \label{figure:hist_all}
\end{figure*}

Turbulence can affect the electron confinement by changing the efficiency of magnetic mirror, as the turbulent flows modulate the field strength along a magnetic field line. Considering this, we evaluate the minimum pitch angles of electrons that are reflected via magnetic mirror for each field line, and we compare the 2D and 3D models. Here, we ignore the pitch-angle scattering.
The outline of the analysis method is as follows.
We pick up a field line that passes through the ALT region in a snapshot data. 
We write the minimum field strength along the field line as $B_{\rm min}$. The electron with the pitch angle larger than the following value will be reflected via magnetic mirror at the location with the field strength of $B_{\rm mir}$:
\begin{equation}
    \theta_{\rm mir} = \arcsin{\left( \sqrt{\frac{B_{\rm min}}{B_{\rm mir}}}\right)}. \label{eq:theta}
\end{equation}
As we are interested in the confinement in the ALT region, we take $B_{\rm mir}$ as the field strength just outside the ALT region (the detailed explanation will be given later). We calculate $\theta_{\rm mir}$ for many different field lines, and we produce the histogram against $\theta_{\rm mir}$ to examine the statistical property.
We expect that the histogram for the 3D model will have a wider distribution because the turbulent flows change $B_{\rm min}$ in a complex way.

We require that the field lines analyzed should pass through a segment of the reconnection outflow that penetrates the ALT region, as we expect that non-thermal electrons are accelerated just around or injected by the reconnection outflow. An example of such a field line is shown in the left panel of Figure~\ref{figure:hist_all}. We calculate $\theta_{\rm mir}$ for 4,000 field lines that meet the requirement. $B_{\rm mir}$ is the average of the magnetic field strength measured at the two intersections of the tracked magnetic field lines and the plane at $y=y_{\rm{btm}}$. For example, $B_{\rm mir}$ for the red field line in Panel (a) of Figure~\ref{figure:hist_all} is obtained by averaging the field strengths at $(x,y)=(1.22L_{0}, 12.4L_{0})$ and $(-1.25L_{0}, 12.4L_{0})$.

The histograms against $\theta_{\rm mir}$ for the 2D and 3D models are compared in the right panel of Figure~\ref{figure:hist_all}. As expected, the 3D model shows a wider distribution than the 2D model. The lower edge of the distribution extends to a smaller $\theta_{\rm mir}$ because the turbulent flows produce the regions with very weak magnetic fields. This result suggests that the turbulent flows can contribute to electron confinement.

\section{Summary and Discussion} \label{sec:summary_discussion}
We performed MHD simulations of a solar flare and investigated the ALT oscillation and the excitation of turbulence in the ALT region. We found that the ALT oscillation can occur in three-dimension in an asymmetric manner even when the reconnection outflow is a quasi-steady laminar flow (Figure \ref{figure:oscillation}) and the ALT region is filled with turbulent flows (Figure~\ref{figure:turbulence_beta}). The ALT oscillation is caused by the asymmetrically vibrating magnetic tuning fork \citep{Takasao2016ApJ,Takahashi2017ApJ}.
The ALT oscillation is found to change the level of turbulence in the arms (Figure \ref{figure:turblent_energy_density}), which indicates a tight relation among the backflow of the reconnection outflow, the ALT oscillation, and turbulence. 

\begin{figure*}
    \centering
    \includegraphics[width=2\columnwidth]{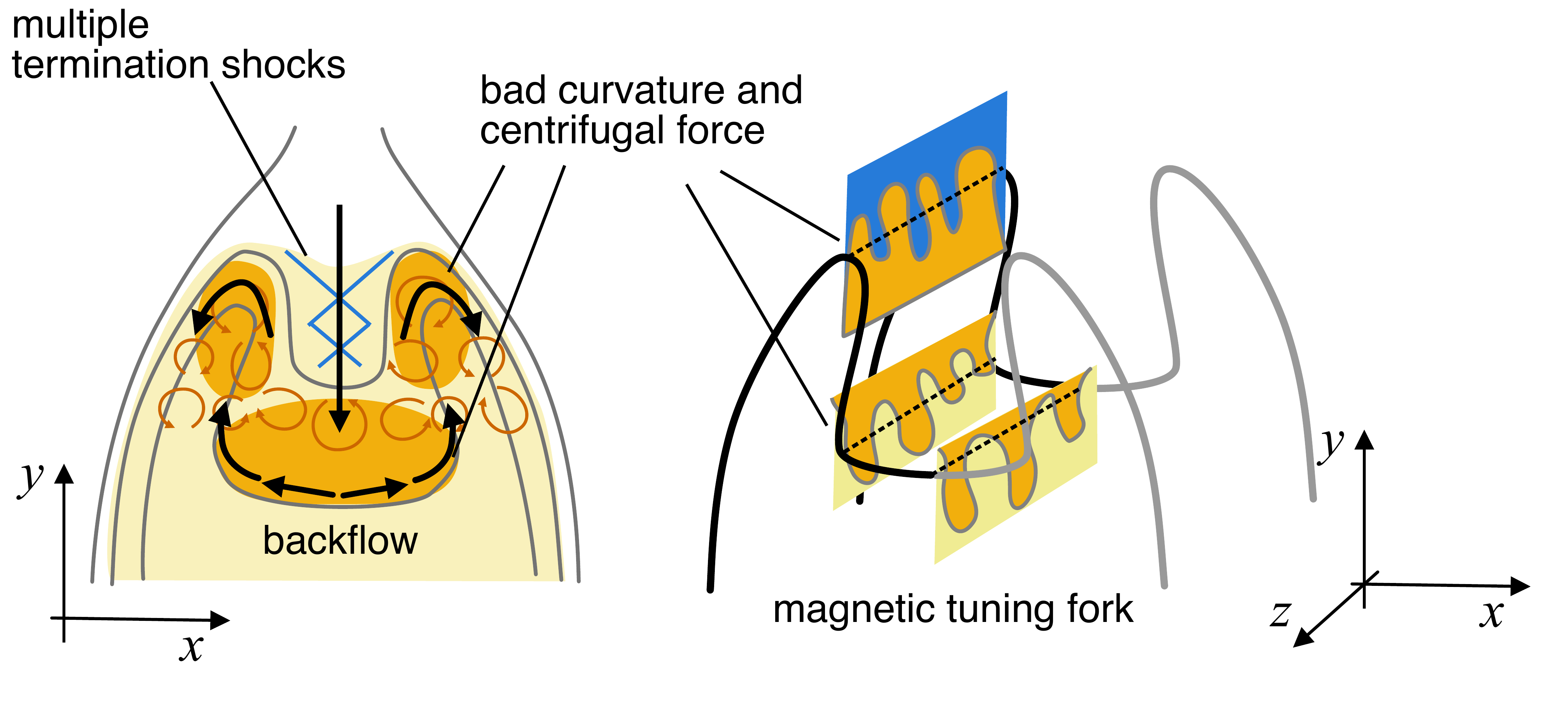}
    \caption{Schematic diagram of instabilities in the ALT regions.}
    \label{figure:summary}
\end{figure*}

Our 3D simulation showed rapid growth of MHD instabilities in the ALT region, particularly in the upper part of the ALT region. Instabilities developing in the arms of the magnetic tuning fork produce turbulent flows which surround the multiple termination shocks.
Figure \ref{figure:summary} summarizes how and where instabilities develop in the ALT region.
The reconnection outflow penetrates the ALT region to produce four bad-curvature regions where pressure-driven instabilities can occur; the bottom end of the reconnection outflow and the two arms of the magnetic tuning fork (Figure \ref{figure:inst}). 
In addition, the transonic backflow moves along the curved magnetic field, which can drive centrifugally driven Rayleigh-Taylor instabilities.
Our simulation demonstrated the development of instabilities in these regions (Figures \ref{figure:turbulence_beta}, \ref{figure:inst} and \ref{figure:turblent_energy_density}), which suggests that the rapidly growing instabilities are combinations of pressure-driven and centrifugally driven modes. 
Regarding the bottom end of the reconnection outflow, Nakamura (2013) also pointed out the formation of finger-like structures via the curvature-driven instability, by performing a 3D simulation of a solar flare. Other related studies will be discussed in Section~\ref{subsec:comparison}.

\subsection{Comparison with observations \\
of the 2017 10 September flare}

\begin{figure*}
    \centering
    \includegraphics[width=15.0cm]{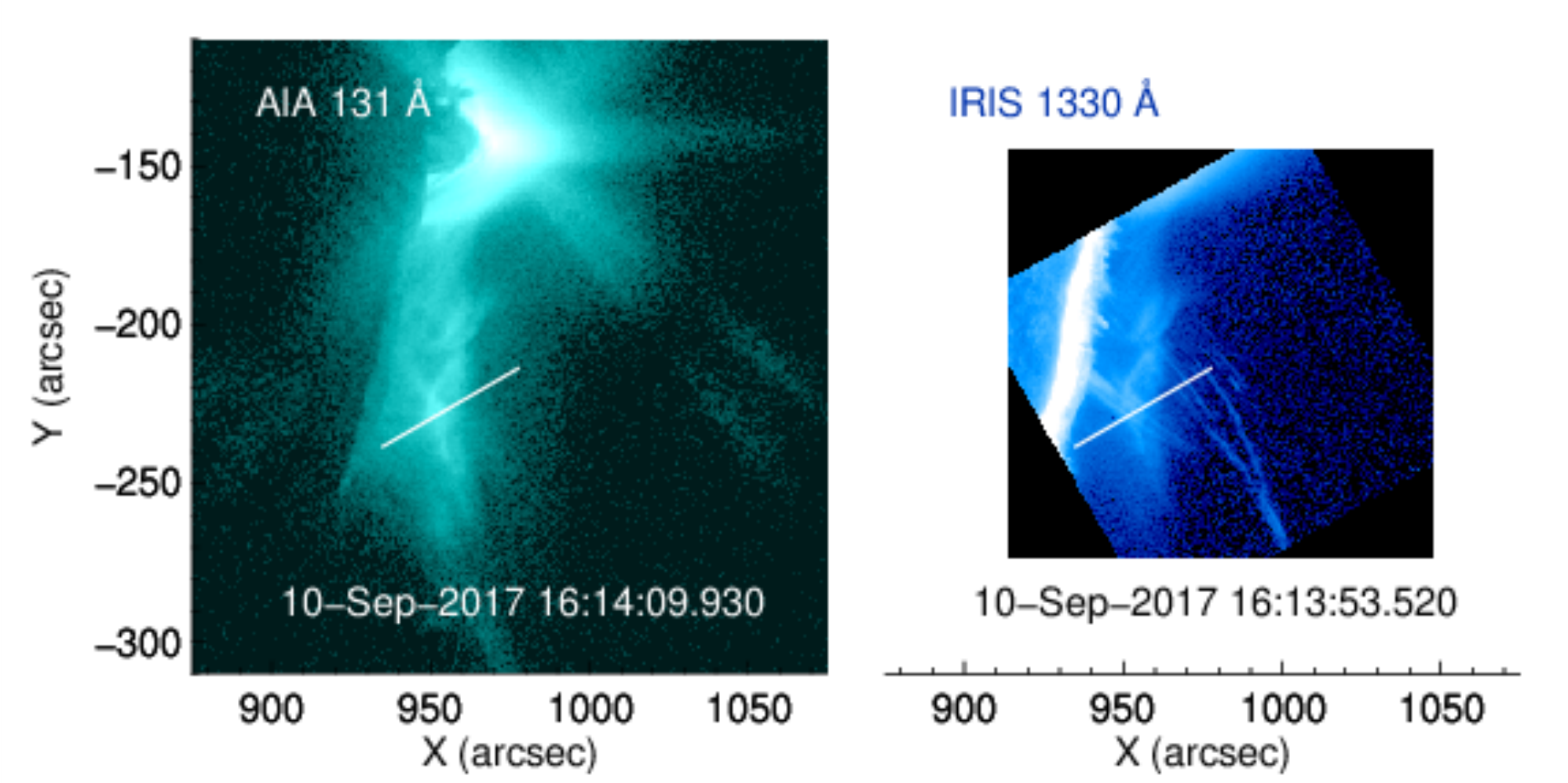}
    \caption{{\it SDO}/AIA 131 \AA\ image (left) and {\it IRIS} 1330 \AA\ slit jaw image (right) showing the location of the slit used in Figure \ref{int_nt_vel.fig}.}
    \label{imaging_summary.fig}
\end{figure*}

To put the modeling in context, we re-examine the {\it IRIS} data presented by \citet{Reeves2020ApJ} of the ALT region in the 2017 10 September flare.  The location of the region of interest is shown in Figure \ref{imaging_summary.fig}.   

\begin{figure*}
    \centering
    \includegraphics[width=15.0cm]{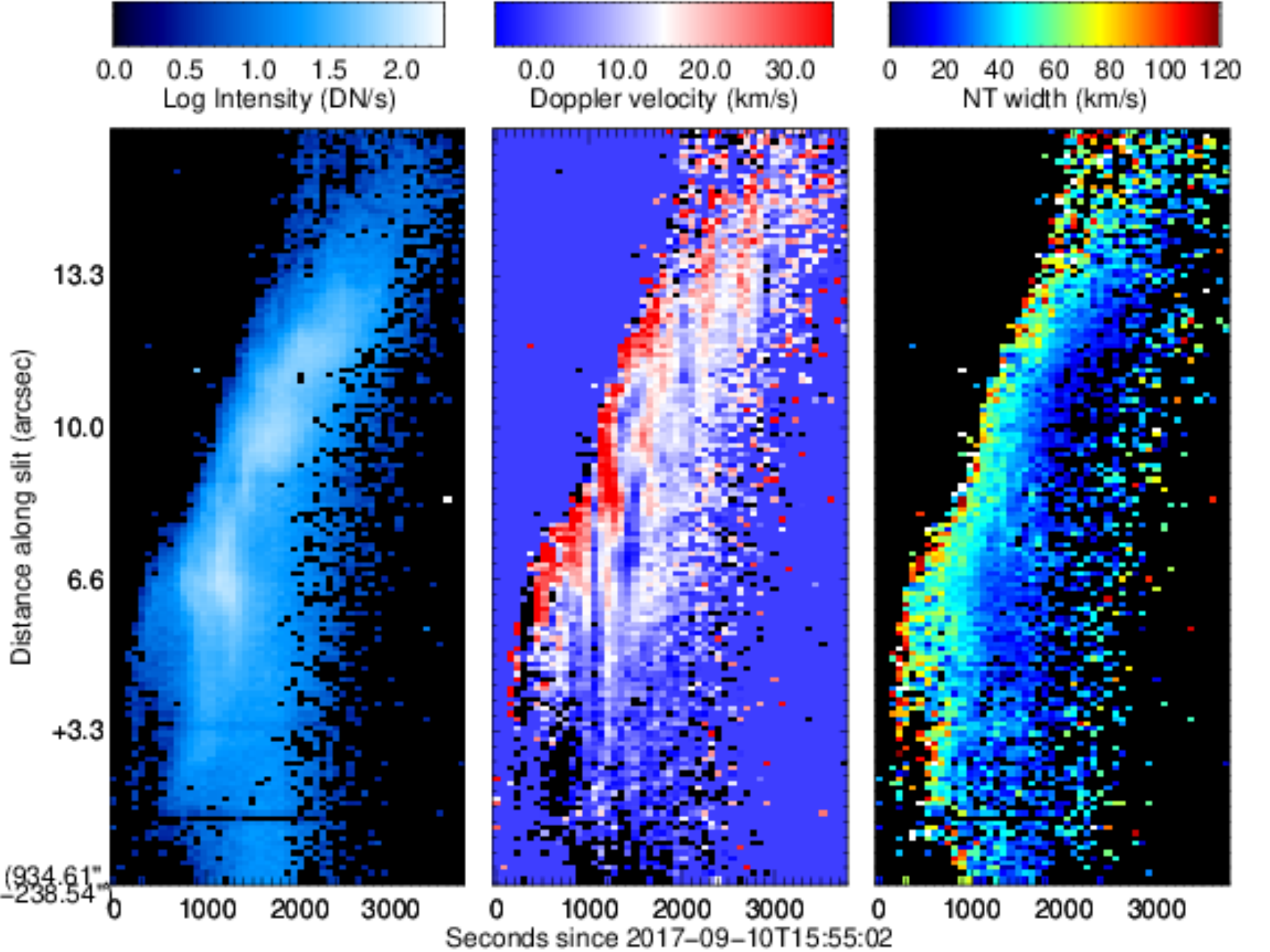}
    \caption{Intensity (left) Doppler velocity (middle) and non-thermal velocity (right) calculated from Gaussian fits of the {\it IRIS} Fe \textsc{xxi} line along the slit shown in Figure \ref{imaging_summary.fig} as a function of time.  Note that the color table for the Doppler velocity is not centered at 0 km s$^{-1}$, but at 15 km s$^{-1}$ in order to visually enhance the oscillations. }
    \label{int_nt_vel.fig}
\end{figure*}

During this event, {\it IRIS} was observing the region to the south of the main flare loops with an eight step raster (see \citet{Reeves2020ApJ} for details of the {\it IRIS} observations).  We calculate Gaussian fits to the Fe \textsc{xxi} line along the slit in the sixth raster position, and show the resulting  intensity, Doppler velocity and non-thermal velocity along the slit as a function of time in Figure \ref{int_nt_vel.fig}.  The Doppler velocity map clearly shows oscillations along the top edge of the structure, indicating that oscillations are present as the ALT region rises, as was found in the model.  The analysis in \citet{Reeves2020ApJ} measured oscillations in the Doppler shifts with periods of $\sim$400 s, but in that analysis, the location of the data analyzed was stationary and not rising with the ALT source.  The rising oscillations found in this re-analysis of the data nevertheless still have similar periods to those found in \citet{Reeves2020ApJ}. We estimate the acoustic Mach number in this region by dividing the velocities indicated by the Doppler shifts by the sound speed, assuming a temperature of 10 MK.  We find numbers in the range of 0.01-0.07, about an order of magnitude smaller than those found in the model and shown in Figure \ref{figure:time_vz}.  The differences may be due to differences in plasma parameters in the model and the observations.

Figure \ref{int_nt_vel.fig} also clearly shows that the highest non-thermal velocity is always at the uppermost visible edge of the ALT region as seen by {\it IRIS}.  This result indicates that there is possibly a large amount of turbulence accompanying the oscillations.

\subsection{Comparison with previous numerical studies}\label{subsec:comparison}
The development of the RT and Richtmyer-Meshkov instabilities around the ALT region have been discussed in previous studies to study the origins of supra-arcade downflows and turbulent flows \citep[e.g.][]{Guo2014ApJ,Shen2022NatAs}. These instabilities only occur in flare simulations when the third direction is included.
In the previous simulations, the instabilities develop at the density interface beneath the reconnecting current sheet (in other words, the bottom end of the reconnection jet).

Our 3D simulation shows the development of instabilities at the density interface at the bottom end of the reconnection jet (Figure~\ref{figure:inst}), as in previous simulations. However, we found that the arms of the magnetic tuning fork develop instabilities more rapidly. 
The growing modes produced at the density interface start to rise well below the termination shock region. Therefore, they have little impact on the termination region in the early phase of the flare. However, as the arms are much closer to the termination shock region, the turbulent flows produced in the arms quickly surround it (Figure~\ref{figure:turbulence_beta}). Therefore, the turbulence produced in the arms has a stronger impact. Figure \ref{figure:hist_all} shows a field line connecting the post-shock region and pre-shock region (see the red line). The field line is produced by the turbulence in the arms. 
We expect that electrons can be accelerated to high energies via multiple energizations (similar to the diffusive shock acceleration) because such a field line helps electrons to cross termination shocks multiple times.
This idea is similar to a picture of the multiple energizations of electrons based on 2D MHD plus kinetic models \citep[e.g.][]{Kong2019ApJ,Li2022ApJ}.

The ALT oscillation in this study is produced by the asymmetrically vibrating magnetic tuning fork. Recurrent ejections of plasmoids \citep[e.g.][]{Kliem2000A&A}, coalescence of plasmoids \citep[e.g.][]{Tajima1987ApJ,Jelinek2017ApJ} or quasi-periodic reconnection \citep[e.g.][]{Craig1991ApJ,Nakariakov2006A&A,McLaughlin2009A&A,Thurgood2017ApJ} are not required to produce the ALT oscillation in our models, although our study does not exclude the possibility that these mechanisms are responsible for some quasi-periodic pulsation (QPP) events \citep[for reviews of QPP, see, e.g.][]{McLaughlin2018SSRv,Zimovets2021SSRv}. Indeed, \citet{Takasao2012ApJ} found recurrent plasmoid ejections in an eruptive solar flare \citep[see also][]{Takasao2016ApJ_plasmoid}. The ALT dynamics caused by plasmoid-mediated reconnection in three-dimension is to be investigated.

\subsection{Growth timescale of the instabilities}
We discuss the growth timescale of the instabilities that develop in the arms of the magnetic tuning fork ($t_{\rm grow}\approx \gamma_{\rm grow}^{-1}$) and compare it to the Alfv\'en timescale of the flare arcade system $t_{\it A,{\rm in}}=L_y/v_{\it A,{\rm in}}$, where $v_{\it A,{\rm in}}$ is the Alfv\'en speed in the reconnection inflow region. Using Equation~(\ref{eq:max_growth_rate_approx}), the growth rate $\gamma_{\rm grow}$ may be approximated as
\begin{eqnarray}
    \gamma_{\rm grow} \sim \frac{c_{\rm s,ALT}}{R_{\rm c}},
\end{eqnarray}
The ratio of the two timescales is estimated as
\begin{eqnarray}
    \frac{t_{\rm grow}}{t_{\it A,{\rm in}}} \sim 0.01\left(\frac{R_{\rm c}/L_y}{0.01}\right) \left( \frac{v_{\it A,{\rm in}}/c_{\rm s,ALT}}{1}\right),\label{eq:ratio_timescale}
\end{eqnarray}
where we assume that the width of arms ($\sim R_{\rm c}$) is approximately 1\% of the system size $L_y$. As the reconnection outflow speed will be similar to $v_{\it A,{\rm in}}$, $v_{\it A,{\rm in}}/c_{\rm s,ALT}$ essentially denotes the acoustic Mach number of the reconnection outflow.
The sound speed in the ALT region will be similar to the Alfv\'en speed if the heat conduction cooling is ignored.
As the growth timescale of turbulence is much shorter than the Alfv\'en timescale of the system, the development of turbulence should be instantaneous in terms of the flare duration. In other words, the ALT region will quickly become turbulent without the injection of the turbulent reconnection outflow.

The above discussion ignores the effect of heat conduction cooling. We discuss the scaling of the timescale ratio in the case with heat conduction.
\citet{Seaton2009ApJ} and \citet{Takasao2016ApJ} show that the acoustic Mach number of the reconnection outflow is larger in the case with heat conduction than that in the adiabatic case, as a result of the conduction cooling. The dependence of the acoustic Mach number is 
\begin{equation}
    \frac{v_{\it A,{\rm in}}}{c_{\rm s,ALT}}\propto \beta^{-2/7}L_y^{-1/7}.\label{eq:mach_hc}
\end{equation}
We note that the conduction cooling also affects the dependence of $R_{\rm c}$. The conduction cooling makes the ALT size $w$ smaller \citep{Takasao2016ApJ}. If we assume that $R_{\rm c}\sim w$, the dependence is
\begin{equation}
    R_{\rm c}\propto \beta^{4/7}L_y^{9/7}.\label{eq:Rc_hc}
\end{equation}
As a result, from Equations (\ref{eq:ratio_timescale}), (\ref{eq:mach_hc}), and (\ref{eq:Rc_hc}), we obtain the following scaling for the case with heat conduction:
\begin{equation}
    \frac{t_{\rm grow}}{t_{\it A,{\rm in}}} \propto \beta^{2/7}L_y^{1/7}.\label{eq:ratio_timescale_hc}
\end{equation}
This result indicates that turbulence via the curvature-driven instabilities will develop more quickly in a flare with a lower plasma $\beta$ (a stronger magnetic field). This plasma $\beta$ dependence mainly comes from the strong $\beta$ dependence of the ALT size (Equation~(\ref{eq:Rc_hc})). Equation (\ref{eq:ratio_timescale_hc}) tells that the flare size $L_y$ has little influence on the growth timescale.

Both the interchange and undular modes can be stabilized if there is a magnetic shear in the ALT region.
However, our model assumes a negligible magnetic shear in the reconnection inflow region. As a result, magnetic shear in the arms of the magnetic tuning fork regions is too small to stabilize the growing modes.

The magnetic shear will also be unimportant for the growth of instabilities in an actual solar flare. 
Let us consider the situation in which magnetic shear is developed on the scale of an active region. In such a case, the magnetic shear in the ALT region will be important if the region contains a significant fraction of the magnetic flux in the active region. However, as we will show below, the ALT region contains a very small magnetic flux. After the onset of a flare, the reconnected field piles up to form a small ALT region. The total flux of the reconnected field in the ALT region $\Phi_{\rm ALT}$ is estimated to be $B_{\rm out} w L_z$, where $B_{\rm out}$ is the field strength in the reconnection outflow, and $L_z$ is the typical length scale of the flare along the neutral line. We also define the nondimensional reconnection rate as $\xi$. Considering the magnetic flux conservation during reconnection, we get $B_{\rm out}\approx \xi B_{\rm in}$, where $B_{\rm in}$ is the field strength in the reconnection inflow region or around the reconnection current sheet. $\Phi_{\rm ALT}$ is then estimated as follows:
\begin{align}
    \Phi_{\rm ALT}&\approx 10^{18}~{\rm Mx} \left( \frac{\xi}{10^{-2}}\right)\left( \frac{w/L_y}{10^{-2}}\right) \nonumber \\
     &\times \left(\frac{B_{\rm in}}{100~{\rm G}}\right)\left( \frac{L_y}{10^{10}~{\rm cm}}\right)\left( \frac{L_z}{10^{10}~{\rm cm}}\right),
\end{align}
where we assume that the size of ALT region is 1\% of the flare size.
The magnetic flux of active regions is typically in the range of $10^{21}\mbox{-}10^{23}$~Mx, which is much larger than the estimated value of $\Phi_{\rm ALT}$. Therefore, magnetic shear in the ALT region should be negligible.
A large magnetic shear may occasionally develop when a significant magnetic shear is stored in the reconnecting field on a very small spatial scale or when plasmoids with a strong guide field are injected into the ALT region. 

\vspace{10pt} 
S.T. was supported by the JSPS KAKENHI grant Nos. JP21H04487, JP22H00134, and JP22K14074.
Numerical computations were in part carried out on Cray XC50 at Center for Computational Astrophysics, National Astronomical Observatory of Japan. This work was achieved through the use of SQUID at the Cybermedia Center, Osaka University.

\bibliography{main_arXiv}{}
\bibliographystyle{aasjournal}

\appendix
\section*{The effects of heat conduction} 
In the main text, we focus on the models without the effect of heat conduction because of the limitation of the computational resources. However, we performed a 3D, lower-resolution simulation with this effect and briefly examined robustness of our results. A second-order, piece-wise linear method (PLM) is used for spatial reconstruction. The simulation includes the following heat conduction flux in the energy equation:
\begin{eqnarray}
    \bm{F_{c}}=-\kappa_{0}T^{5/2}\nabla_{\parallel}T,
\end{eqnarray}
where $\kappa_0 = 8.2\frac{k_{B}}{m_{\rm ion}}\frac{L_{0}c_{\rm iso, 0}\rho_{0}}{T_{0}^{5/2}} =  9.75\times 10^{-7}$ in cgs units, and $\nabla_{\parallel}$ denotes the gradient parallel to the magnetic field.
The calculation domain is $-7.5L_{0} \leq x \leq 7.5L_{0}$, $0\leq y\leq 20L_{0}$, and $-0.5L_{0}\leq z \leq 0.5L_{0}$ (the length in the $z$ direction is shorter than that of the simulation in the main article). This domain is resolved by a $450 \times 600 \times 30$ grid. The resolution for one direction is two times lower than that of the simulation in the main article.
To update the energy equation, the MHD and heat conduction parts are solved in an operator split manner. For the heat conduction part, we used the second-order  Super TimeStepping method \citep{Meyer2012MNRAS}.

\begin{figure*}
    \centering
    \includegraphics[width=18.0cm]{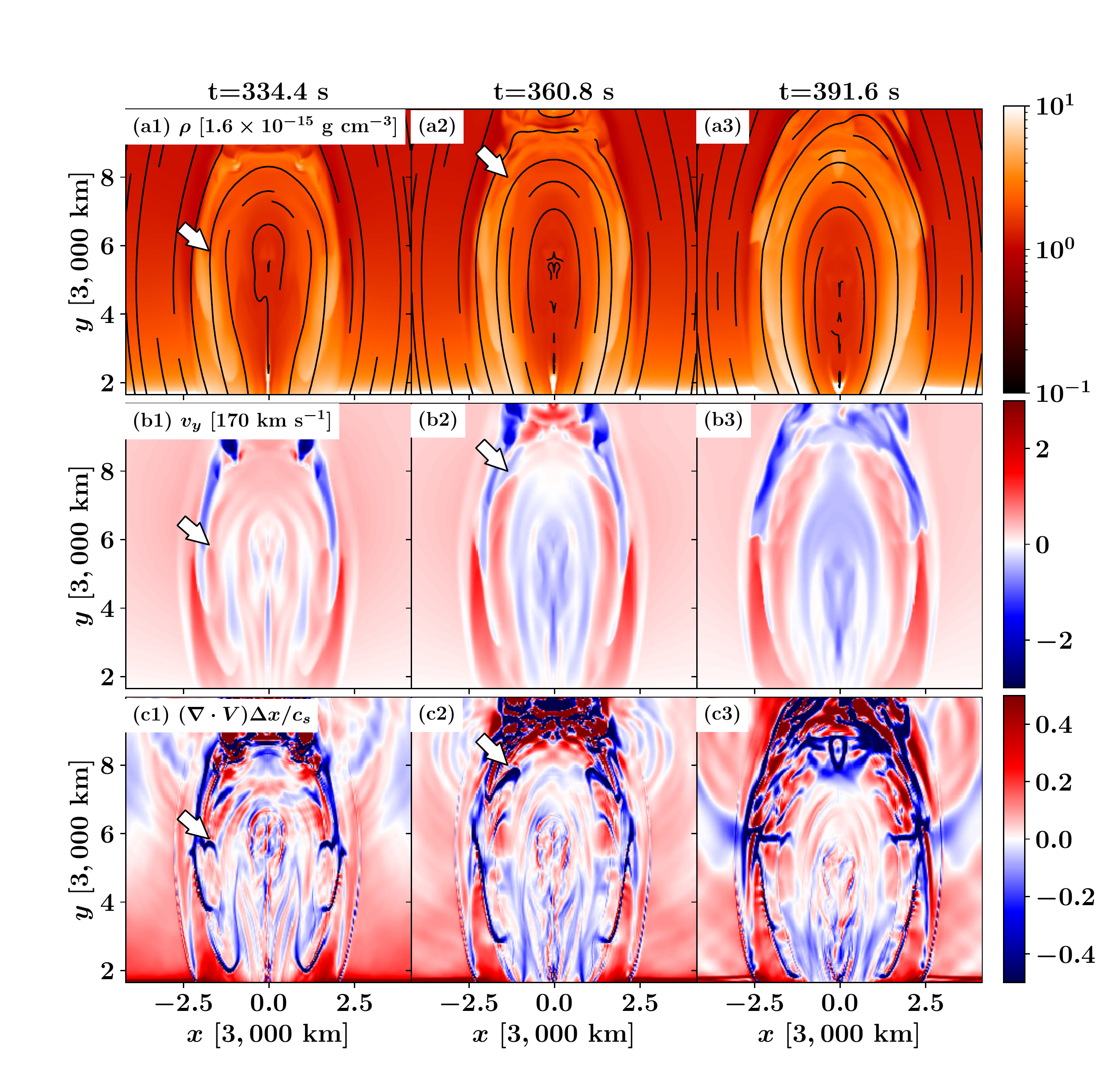}
    \caption{The 2D snapshots of a simulation with heat conduction. The panels in the top, middle, and bottom rows show the density $\rho$, the vertical velocity $v_y$, and the normalized divergence of the velocity field, respectively. The solid lines in the density map indicate the projected magnetic field structure. The white arrows indicate the top of chromospheric evaporation.}
    \label{figure:hc}
\end{figure*}

The overall flare loop structure is largely affected by heat conduction, but it has little effect on the ALT dynamics. The growth of the flare loop is shown in Figure \ref{figure:hc}. 
The chromospheric evaporation flows supply hot and dense plasma to the flare loop.
As the evaporation flows move upward at a transonic or supersonic speed, strong compression occurs at their heads. Considering the property of the evaporation flows, we can identify the head of an evaporation flow by looking at $\rho$, $v_y$, and the normalized $\nabla \cdot \bm{v}$. The head is indicated by the white arrows in the figure. Note that the counter-moving evaporation flows have collided with each other to form a high-density region around at $t=391.6$~s \citep[see also][]{Takasao2015ApJ}. Although the evaporation flows climb up the flare loops, the evaporation flows do not reach the ALT region. Namely, $dy_{\rm top}/dt > v_{\rm evap}$, where $v_{\rm evap}$ is the speed of the evaporation flows. Therefore, the evaporation flows have little influence on the ALT region. We note that the relative magnitude between $dy_{\rm top}/dt$ and $v_{\rm evap}$ will depend on the reconnection process. The two speeds could be estimated from the rise speed of the hard X-ray source and the rise speed of the soft X-ray loops \citep[e.g.][]{Shimizu2008ApJ}.

\begin{figure*}
    \centering
    \includegraphics[width=15.0cm]{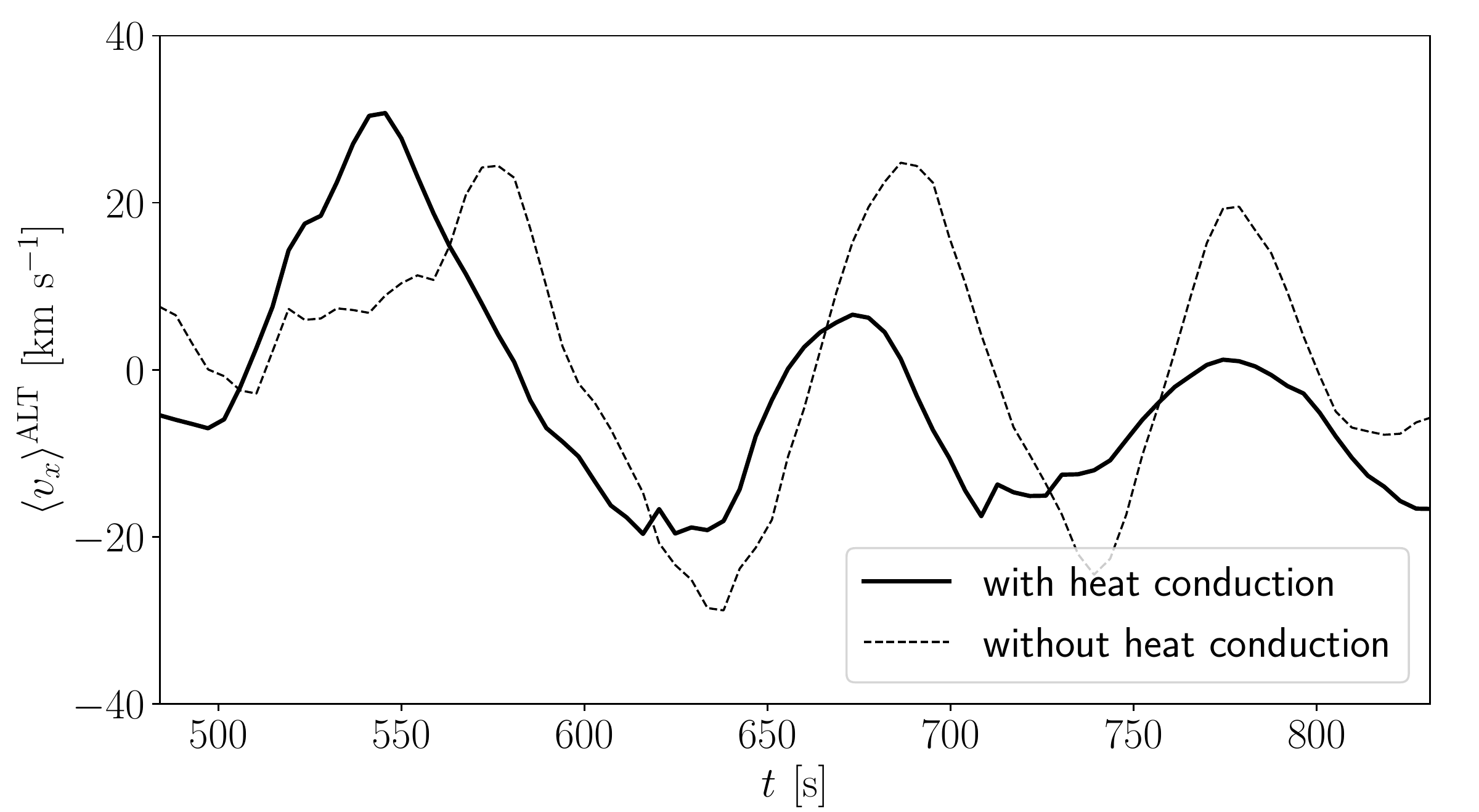}
\caption{Time evolution of the emission-measure-weighted horizontal velocity, $\langle v_x\rangle^{\rm ALT}(t)$. The solid line denotes the result of the 3D model with heat conduction, and the dashed line indicates the result of the 3D model without heat conduction (i.e., the model introduced in the main text). The definition of $\langle v_x\rangle^{\rm ALT}(t)$ is the same as Figure \ref{figure:oscillation}.}
 \label{figure:oscillation2}
\end{figure*}

The ALT oscillation is also found in this model. Figure \ref{figure:oscillation2} is the same as Figure \ref{figure:oscillation} but for the model with the effect of heat conduction. The oscillation period is approximately 100 to 130 s, and the maximum velocity amplitude is approximately $20~{\rm km~s^{-1}}$, which are similar to those of the case without heat conduction. However, the damping of the velocity amplitude seems to be more prominent in this case. The damping may be promoted by heat conduction cooling, but the low spatial resolution may also be a reason. We need to study the resolution dependence for better understanding of the ALT oscillation.

\begin{figure*}
    \centering
    \includegraphics[width=18.0cm]{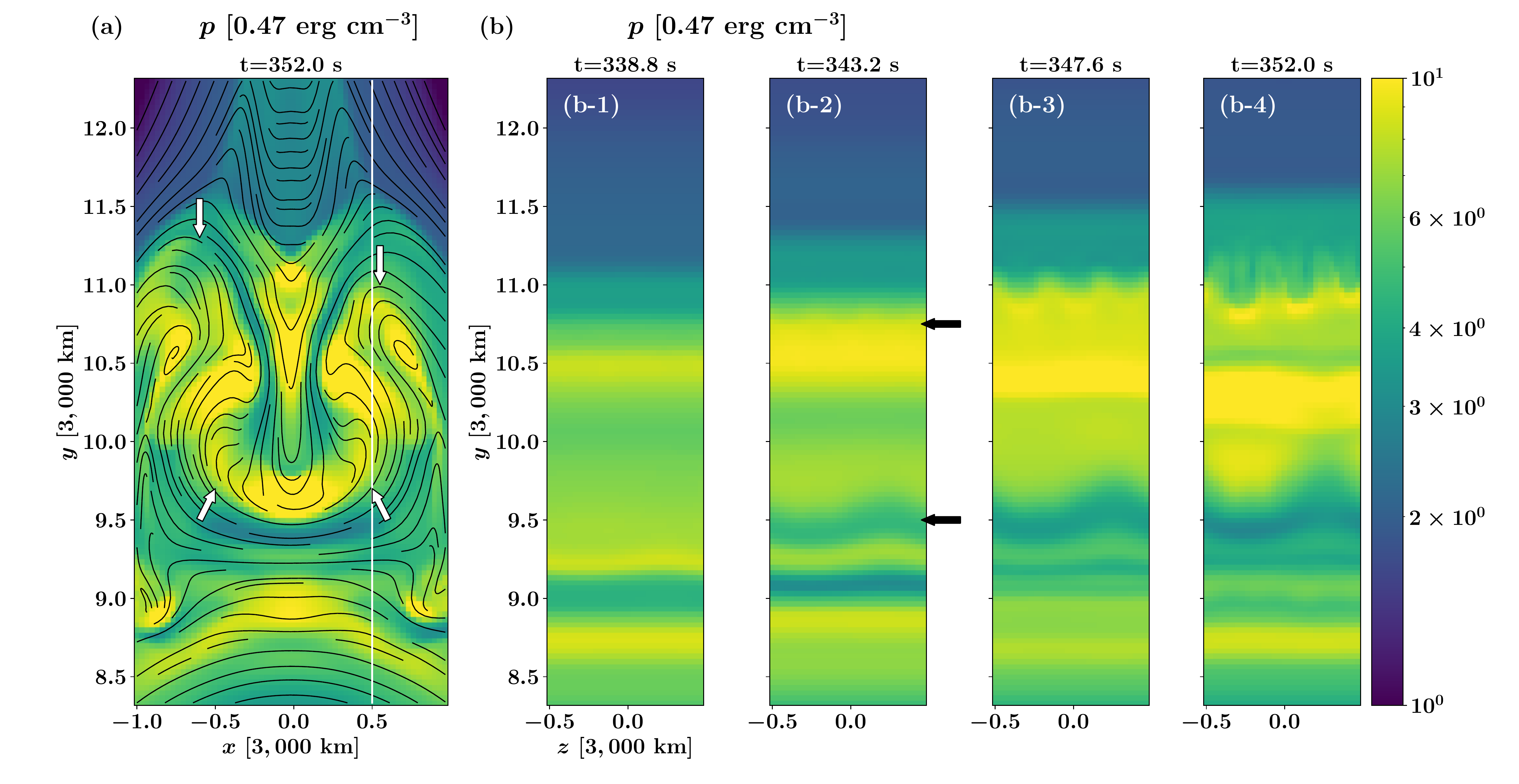}
    \caption{Panel (a) shows the pressure distribution around the ALT region, where bad-curvature regions are indicated by white arrows. Black lines represent the magnetic field lines projected onto this plane. Panels (b-1)-(b-4) display the development of instabilities. The color shows the pressure in the $yz$ plane indicated by the white line in Panel (a). The black arrows indicate the height where the instability occurs.}
 \label{figure:pressure_hc}
\end{figure*}

\begin{figure}
    \centering
    \includegraphics[width=\columnwidth]{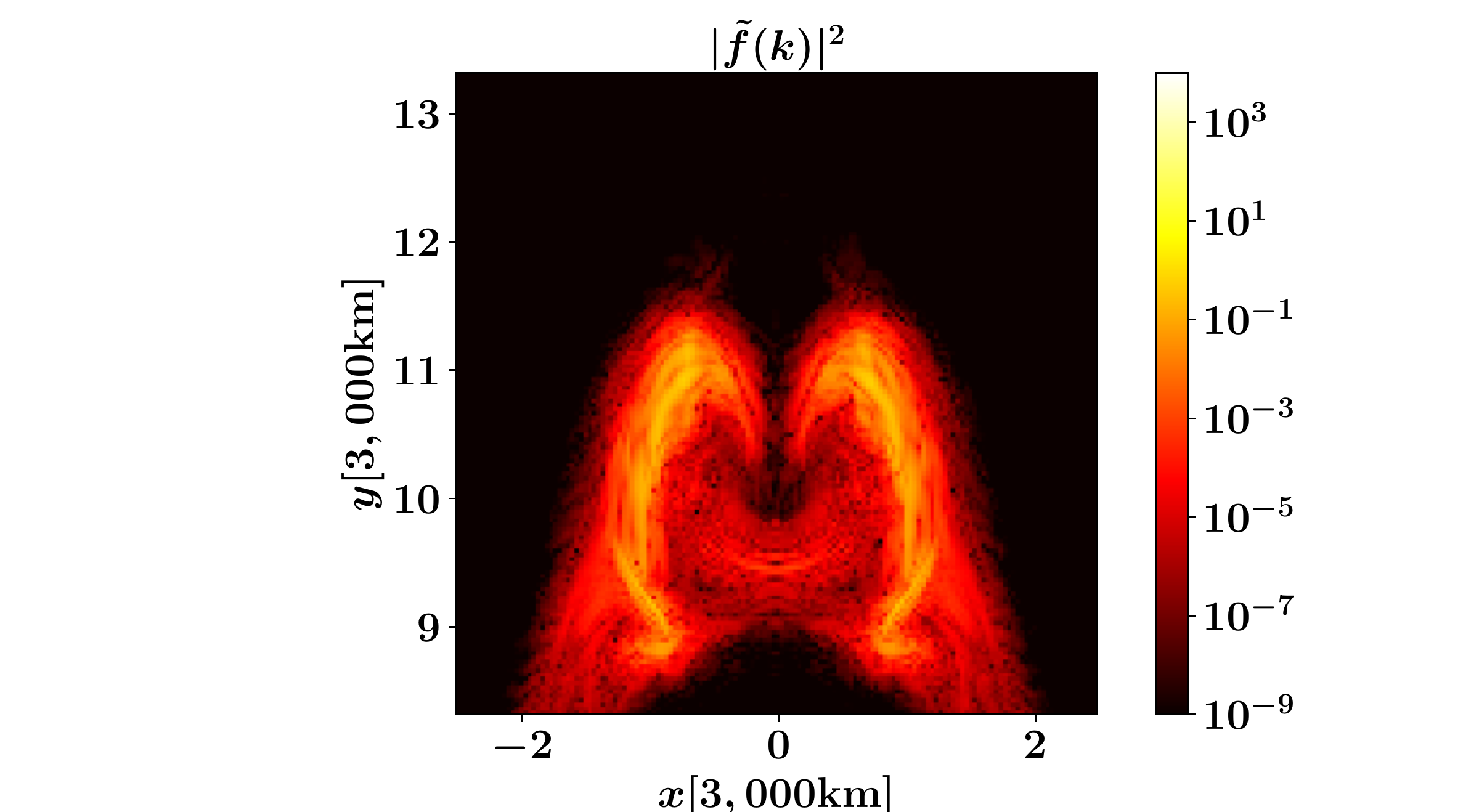}
    \caption{The spatial distribution of the Fourier power of the density fluctuation for the 3D model with the effect of heat conduction. The power corresponding to a wave number of $k=2.3\ [\rm{cell\ size^{-1}}]$ is shown. The data is taken at $t=352\ \rm{s}$.}
 \label{figure:power_HC}
\end{figure}

Figure \ref{figure:pressure_hc} displays the development of the instabilities in the ALT region. As discussed in Section~\ref{sec:instability}, we find the development of the instabilities in the regions with bad-curvature. Figure \ref{figure:power_HC} indicates the strong Fourier power in the arms of the magnetic tuning fork, which demonstrates that instabilities grow particularly in the arms even with the effect of heat conduction.

\begin{figure*}
    \centering
    \includegraphics[width=18.0cm]{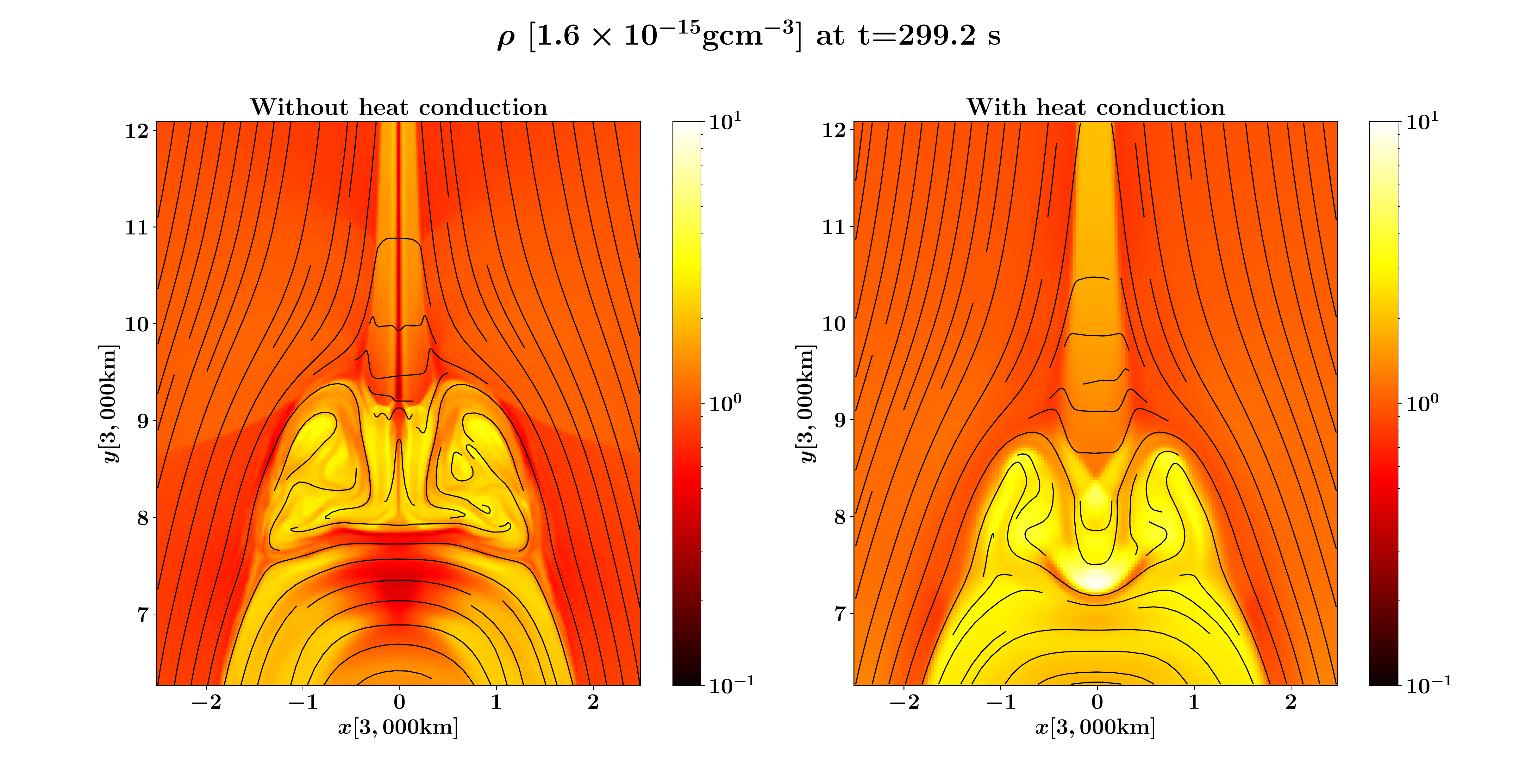}
    \caption{The comparison of the size of ALT region between a model without heat conduction and a model with heat conduction. The color shows the mass density, and the black lines show the magnetic field lines. The time of snapshot is 299.2~$\rm{s}$.}
 \label{figure:ALTcompare}
\end{figure*}

We note that the instabilities grow more rapidly in the case with heat conduction than in the model without it. To find the reasons, we compare the ALT regions of the two models at the same time (Figure \ref{figure:ALTcompare}). With heat conduction, the ALT size is smaller and the density is larger because of the conduction cooling \citep[see also][]{Takasao2016ApJ}. The smaller ALT size results in the smaller curvature radius. The larger density contrast leads to a larger Atwood number for the Rayleigh-Taylor instabilities. We consider that these two effects increase the growth rate.

\end{document}